\documentclass[fleqn,usenatbib]{mnras}

\usepackage{newtxtext,newtxmath}

\usepackage[T1]{fontenc}

\DeclareRobustCommand{\VAN}[3]{#2}
\let\VANthebibliography\thebibliography
\def\thebibliography{\DeclareRobustCommand{\VAN}[3]{##3}\VANthebibliography}

\def\be{\begin{equation}}
\def\ee{\end{equation}}

\usepackage{graphicx}	
\usepackage{amsmath}	
\usepackage{multirow}
\usepackage{hyperref}

\usepackage{CJKutf8}

\newcommand{\eddurl}{https://edd.ifa.hawaii.edu/}

\title[The SDSS PV Catalogue]{The Sloan Digital Sky Survey Peculiar Velocity Catalogue}

\author[Cullan Howlett et al.]{
Cullan Howlett,$^{1}$\thanks{E-mail: c.howlett@uq.edu.au}
Khaled Said$^{1}$,
John R. Lucey$^{2}$,
Matthew Colless$^{3}$,
Fei Qin$^{4}$,
Yan Lai$^{1}$,\newauthor
R. Brent Tully$^{5}$,
Tamara M. Davis$^{1}$
\\
$^{1}$School of Mathematics and Physics, The University of Queensland, Brisbane, QLD 4072, Australia. \\
$^{2}$ Centre for Extragalactic Astronomy, Durham University, Durham DH1 3LE, United Kingdom\\
$^{3}$Research School of Astronomy and Astrophysics, Australian National University, Canberra, ACT 2611, Australia.\\
$^{4}$Korea Astronomy and Space Science Institute, Yuseong-gu, Daedeok-daero 776, Daejeon 34055, Korea. \\
$^{5}$Institute for Astronomy, 2680 Woodlawn Drive, Honolulu, HI 96822, USA.
}

\date{Accepted XXX. Received YYY; in original form ZZZ}

\pubyear{2020}

\begin{document}
\label{firstpage}
\pagerange{\pageref{firstpage}--\pageref{lastpage}}
\maketitle

\begin{abstract}
We present a new catalogue of distances and peculiar velocities (PVs) of $34,059$ early-type galaxies derived from Fundamental Plane (FP) measurements using data from the Sloan Digital Sky Survey (SDSS). This $7016\,\mathrm{deg}^{2}$ homogeneous sample comprises the largest set of peculiar velocities produced to date and extends the reach of PV surveys up to a redshift limit of $z=0.1$.  Our SDSS-based FP distance measurements have a mean uncertainty of 23\%. Alongside the data, we produce an ensemble of 2,048 mock galaxy catalogues that reproduce the data selection function, and are used to validate our fitting pipelines and check for systematic errors. We uncover a significant trend between group richness and mean surface brightness within the sample, which may hint at an environmental dependence within the FP or the presence of unresolved systematics, and can result in biased peculiar velocities. This is removed using multiple FP fits as function of group richness, a procedure made tractable through a new analytic derivation for the integral of a 3D Gaussian over non-trivial limits. Our catalogue is calibrated to the zero-point of the CosmicFlows-III sample with an uncertainty of $0.004$ dex (not including cosmic variance or the error within CosmicFlows-III itself), which is validated using independent cross-checks with the predicted zero-point from the 2M++ reconstruction of our local velocity field. Finally, as an example of what is possible with our new catalogue, we obtain preliminary bulk flow measurements up to a depth of $135\,h^{-1}\mathrm{Mpc}$. We find a slightly larger-than-expected bulk flow at high redshift, although this could be caused by the presence of the Shapley supercluster which lies outside the SDSS PV footprint.
\end{abstract}

\begin{keywords}
catalogues -- galaxies: statistics -- galaxies: elliptical and lenticular -- galaxies: fundamental parameters -- galaxies: distances and redshifts -- cosmology: observations
\end{keywords}



\section{Introduction}
\label{sec:introduction}

Galaxies are receding from us due to the expansion of the Universe. The observed relation between galaxy recession velocity and comoving distance is called the Hubble–Lema\^itre law. The variation of any galaxy's observed velocity from its recession is called galaxy peculiar velocity. The main cause of peculiar velocities are the gravitational attraction of the growing large-scale structures (LSS). Hence, robust and accurate measurements of local peculiar velocities are essential for inferring the Hubble-Lema\^itre law and additionally allow for cosmography and precise cosmological studies of gravity in the local Universe.

The peculiar velocity of a galaxy can be derived if one can independently measure both its distance and redshift. Several distance indicators have been developed that enable the mapping of peculiar velocities in the local Universe. Well-known examples include Cepheid variable stars \citep{Leavitt1912}, the tip of the red giant branch \citep{Lee1993}, Type Ia supernovae \citep{Phillips1993}, surface brightness fluctuations \citep{Tonry1988}, the Tully-Fisher relation (TF; \citealt{Tully1977}), the Fundamental Plane (FP; \citealt{Djorgovski1987,Dressler1987}), and gravitational waves \citep{Holz2005}.

Each distance indicator has its own limitations; TF and FP galaxies are relatively abundant and easy to measure, and so far are the only indicators that have been used to derive distances for thousands of galaxies. However, this comes at the cost of large intrinsic scatter in their empirical relationships and so large distance uncertainties.

The current largest individual peculiar velocity samples are the Cosmicflows-IV Tully-Fisher catalogue (CF4-TF; \citealt{Kourkchi2020}) containing $\sim$9,800 objects, and the FP-based 6-degree Field Galaxy Survey peculiar velocity sample (6dFGSv; \citealt{Springob2014}), containing $\sim$8,800 objects. In addition to these individual catalogues, the Cosmicflows project \citep{Tully2008,Tully2013,Tully2016} aims to provide a single comprehensive collection of distance measurements from all of the aforementioned distance indicators. The most recently released iteration, Cosmicflows-III \citep{Tully2016}, contains almost 18,000 galaxies with distance measurements, with progress towards enlarging this substantially (as evident by the recent release of the CF4-TF subsample referenced above). Most previous efforts, including the above cases, have focussed on the $z<0.05$ universe, as this nearby regime is where the FP and TF methods, with large uncertainties that increase with distance, are most useful.

Peculiar velocity catalogues have formed the backbone for many science applications over the years. In the 1990s they were primarily used to constrain $\Omega_m$ and linear galaxy bias $b$ \citep{Willick1997,Sigad1998}. However, as discussed comprehensively in \cite{DavisM1996}, there were inconsistencies between the velocity fields measured by peculiar velocity surveys and predicted by redshift surveys; arising from a combination of sparseness in the redshift surveys at high redshift, angular incompleteness in the peculiar velocity surveys, and potential systematics in the estimation and calibration of the peculiar velocities. More recently, the advent of large surveys of nearby galaxies, such as the 2MASS Redshift Survey \citep{Huchra2012}, Sloan Digital Sky Survey (SDSS; \citealt{York2000}), 6dF Galaxy Survey \citep{Jones2004}, and the Arecibo Legacy Fast ALFA Survey \citep{Giovanelli2005} have enabled the creation of large, homogeneous redshift and PV samples. These samples have demonstrated better consistency (e.g., \citealt{DavisM2011}).

Studies using peculiar velocities have hence seen a resurgence, including in the areas of cosmography \citep{Springob2014,Tully2014,Graziani2019}, measurements of the bulk flow and low-order velocity moments \citep{Watkins2009, Feldman2010, Nusser2011, Ma2013, Scrimgeour2016, Qin2018, Qin2019a,Qin2021,Qin2021b}; testing $\Lambda$CDM and General Relativity via the velocity correlation function or power spectrum \citep{Johnson2014,Huterer2017,Howlett2017c,Adams2017,Howlett2019,Qin2019b,Adams2020}; and fitting cosmological parameters and the external tidal field using reconstructions of the velocity field from galaxy redshifts \citep{Carrick2015,Boruah2020,Said2020,Lilow2021,Stahl2021}. Furthermore, in the era of the Hubble tension (e.g., \citealt{Verde2019}), the importance of robust and accurate peculiar velocity measurements for correcting low-redshift distance measurements from Type~Ia supernovae and gravitational waves has come to the forefront \citep{Scolnic2014,Guidorzi2017,Howlett2020,Boruah2021}.

In this paper we capitalise on these previous efforts, particularly 6dFGSv and its predecessors EFAR \citep{Wegner1996}, SMAC \citep{Hudson1999} and ENEAR \citep{DaCosta2000}, to provide distances and peculiar velocities for more than 30,000 early-type galaxies using the FP relation and data from SDSS. This catalogue is $\sim$3$\times$ larger than either the 6dFGSv or the CF4-TF catalogue, and also larger than the full ensemble of distances in Cosmicflows-III. The size of this catalogue is in large part due to our inclusion of galaxies up to $z=0.1$, which extends the reach of our new measurements beyond those typically produced with the FP or TF relationships and into a region of the Universe that will likely be of increasing interest over the coming years. Compared to previous measurements, our new catalogue is limited to a relatively small sky area ($\sim$7,000\,deg$^2$), but has substantially higher number density than other catalogues in the same redshift regime. 

As a by-product of this work (specifically, of testing our pipeline for converting the SDSS measurements to peculiar velocities), we also provide a suite of 2,048 highly realistic and well-calibrated simulations of the SDSS peculiar velocity catalogue. In combination with the data, these add significant value for future uses of this work, including for cosmological measurements, characterising the local velocity field, and in understanding potential sources of statistical and systematic errors. The PV catalogue, input data, simulations, and associated data products are all publicly available (see Data Availability).

This paper is organised as follows. We describe the data in Section~2. In Section~3 we present the mock catalogues. Fitting the FP parameters is presented in Section~4, while fitting the distances is presented in Section~5. In Section~6 we provide an example use of the data and simulations by measuring the bulk flow of the catalogue. We conclude in Section~7. Finally, the Appendices provide information on a useful straight-line fitting package we have created for data with errors on $x$ and $y$, followed by some mathematical results simplifying the application of a 3D Gaussian model for the FP. Unless otherwise stated, in this paper we assume a flat $\Lambda$CDM cosmological model with $\Omega_{m}=0.31$ and $H_0=100h$\,km\,s$^{-1}$\,Mpc$^{-1}$. All uses of `$\log$' and should be taken to mean logarithms taken to the base 10.

\section{Sloan Digital Sky Survey data}
\label{sec:data}

\subsection{Primary selection criteria}
\label{sec:selection}

The SDSS peculiar velocity catalogue presented here is based on FP data presented in \cite{Said2020}, which was in turn extracted from imaging and spectra provided with the Sloan Digital Sky Survey Data Release 14 \citep[DR14;][]{Abolfathi2018}.

As well as the baseline selection imposed on the SDSS data (which is complete for extended sources at $r$-band Petrosian magnitudes less than 17.7), we apply a number of additional selection criteria. These again align closely with \cite{Said2020}, although there are some differences, and so our full set of criteria are described below. These criteria are designed to isolate dispersion-supported, early-type galaxies with no evidence of recent star formation and robustly measured $z<0.1$ redshifts. We also make use of existing H$\alpha$ measurements and velocity dispersions from the Portsmouth groups DR8 and DR12 catalogues \citep{Thomas2013}\footnote{Available here: \url{https://www.sdss.org/dr12/spectro/galaxy_portsmouth/}}. By cross-matching with the Portsmouth DR8 and DR12 catalogues we have stellar velocity dispersions $\sigma$, uncorrected for fibre aperture effects. The numbers of galaxies remaining after each successive selection criterion are summarised in Table~\ref{tab:selectioncriteria}. Our selection criteria are:
\begin{enumerate}
    \item Objects spectroscopically classified as GALAXY with redshift warning flag $\mathrm{ZWARN}=0$ (i.e.\ no known problems);
    \item de~Vaucouleurs magnitude in the SDSS $r$-band in the range $10.0\,\leq\,m_r\,\leq\,17.0$;
    \item CMB-frame redshift range 0.0033\,$\leq$\,$z$\,$\leq$\,0.1;
    \item likelihood of the surface brightness profile fit with the de~Vaucouleurs model is higher than with the exponential model in both $i$ and $r$ bands;
    \item concentration index $r_{90}/r_{50}$ in $i$ and $r$ bands greater than 2.5;
    \item axial ratio $b/a$ greater than 0.3 (relatively face-on galaxies) in $i$ and $r$ bands;
    \item within the SDSS North Galactic Cap contiguous area;
    \item H$\alpha$ measurements from the Portsmouth DR8 or DR12 catalogues with equivalent width $< 1\text{\normalfont\AA}$;
    \item $g-r$ colours in the range $0.63 - 0.02(M_r + 20) \leq g-r \leq 1.03 - 0.02(M_r + 20)$ (inspired by \citealt{Masters2008}) and
    \item stellar velocity dispersion greater than 70\,km\,s$^{-1}$ (the SDSS spectral resolution limit) and less than 420\,km\,s$^{-1}$ (which removes objects with spurious measurements).
\end{enumerate}

Table~\ref{tab:selectioncriteria} shows that the selection criteria that remove the largest proportions of galaxies are the faint magnitude limit, the redshift range, the requirement that a de~Vaucouleurs profile is more likely than an exponential profile, and the lack of detectable H$\alpha$. The magnitude limit ensures that the later photometric properties we select on are robust and trustworthy, while the maximum redshift removes objects that, even if they were on the FP, would have such large distance errors and such high chance of introducing systematic bias (which is exacerbated as the redshift increases), that we do not feel they are worth including. The cuts on profile likelihood and H$\alpha$ are crucial and required to ensure that the galaxies in our sample are a clean and representative set of early-type galaxies that are \textit{expected} to be on the FP. As we will show in the next section, our selection criteria are largely, but not perfectly, successful in recovering low-$z$ early-type galaxies, and we implement a further visual classification (see Section~\ref{sec:spirals}) to remove remaining, interloping spirals. 

\begin{figure*}
\centering
\includegraphics[width=1.0\textwidth, trim=0pt 0pt 0pt 0pt, clip]{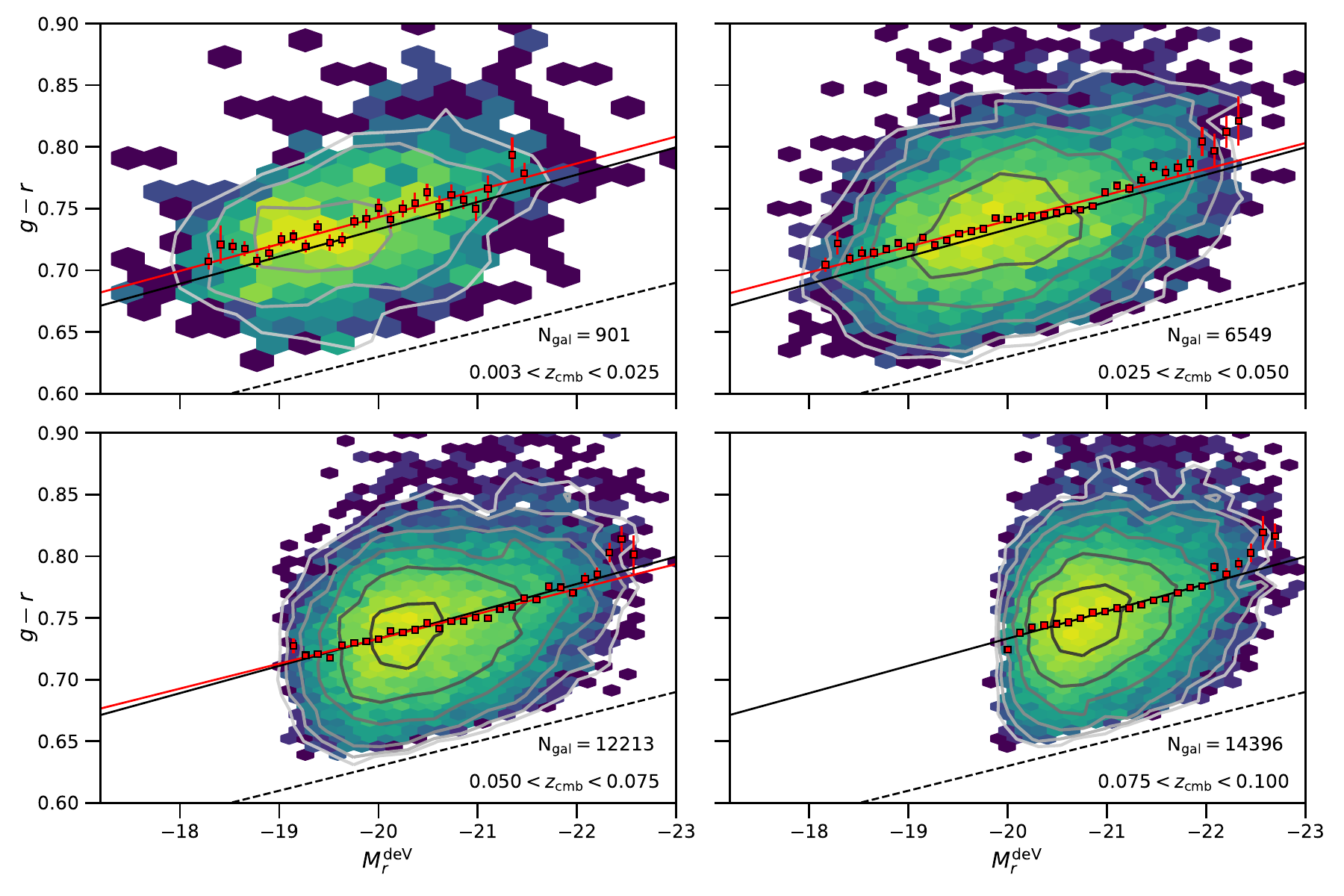}
  \caption{Extinction and $k$-corrected $g-r$ colour vs., $r$-band absolute magnitude for the SDSS peculiar velocity catalogue in four different redshift bins. Hexbins and contours show the density of galaxies in the colour-magnitude space, while red points are averages in bins of absolute magnitude. Red lines show the fits to these points in the lowest three redshift bins, while the black line is the fit to the highest redshift bin, replicated in each of the panels. The dashed-line shows the colour-cut we apply to isolate red objects for our catalogue. The typical galaxy colours are extremely uniform across the entire redshift range of our sample, with comparable slope between the red and black lines in all redshift bins, and a maximum difference of $\sim 0.01$ mags. Also of note is the clear impact of our apparent magnitude limit at higher redshifts, a selection effect that is accounted for when fitting the FP and recovering peculiar velocities.}
  \label{fig:colourmag}
\end{figure*}

We have critically evaluated all the cuts listed above and ensured they are extremely uniform in terms of the redshift distribution of objects they remove, especially the H$\alpha$ EW cut, the $g-r$ colour cut, and the visual inspection described in Section~\ref{sec:spirals}. An example (the colour-magnitude diagram) is shown in Fig.~\ref{fig:colourmag}. This is relevant because, in the course of this work, we found that small changes (i.e., 0.05 mag) in the galaxy properties with redshift can cause large (several hundred km\,s$^{-1}$) biases in the average peculiar velocity of objects at the high redshift end of our sample. Previous works, limited to lower redshift, have rightly assumed or demonstrated that such changes are small enough not to cause a bias in their peculiar velocities. However, photometric or spectroscopic systematics that would have previously been negligible are no longer so when we approach $z=0.1$.

\begin{table}
\centering
\caption{Selection criteria applied to SDSS data to create the SDSS Peculiar Velocity catalogue. Each row summarises a different selection criterion, references the section of the text where it is described, and gives the number of remaining galaxies in the sample \textit{after} this selection has been applied.}
\begin{tabular}{llr} \hline\hline
Selection & Ref. & \# remaining \\ \hline
GALAXY with ZWARN $=0$ & \S\ref{sec:selection}(i) & $403,789$ \\
Magnitude in range $10.0\,\leq\,m_r\,\leq\,17.0$ & \S\ref{sec:selection}(ii) & $287,974$ \\
Redshift range 0.0033\,$\leq$\,$z$\,$\leq$\,0.1 & \S\ref{sec:selection}(iii) & $242,419$ \\
de~Vaucouleurs profile & \S\ref{sec:selection}(iv) & $124,050$ \\
Concentration index $r_{90}/r_{50}\,>\,2.5$ & \S\ref{sec:selection}(v) & $109,614$ \\
Axial ratio $b/a\,>\,0.3$ & \S\ref{sec:selection}(vi) & $102,747$ \\
Within the contiguous NGC area & \S\ref{sec:selection}(vii) & $87,002$ \\
H$\alpha$ EW $< 1\text{\normalfont\AA}$ & \S\ref{sec:selection}(viii) & $45,716$ \\
$g-r$ colour cut & \S\ref{sec:selection}(ix) & $43,226$ \\
Velocity dispersion cut & \S\ref{sec:selection}(x) & $42,170$ \\
No spirals or visual inspection rejects & \S\ref{sec:spirals} & $34,562$ \\
No FP outliers & \S\ref{sec:FPoutliers} & $34,059$ \\
\end{tabular}
\label{tab:selectioncriteria}
\end{table}

\subsection{The FP input parameters}
\label{sec:fpinputs}

Following our primary selection criteria presented in the previous section, we have a number of SDSS photometric and spectroscopic measurements available for each of our galaxies. These include the $r$-band scale radius, $r^\mathrm{deV}$, magnitude $m^{\mathrm{deV}}_{r}$, and axial ratio $(b/a)$, all obtained from de~Vaucouleurs profile fits to the SDSS photometry. We also have the de~Vaucouleurs $g$-band magnitude, from which we construct $g-r$ colours, and $r$-band Galactic extinctions $A_{r}$ \citep{Schlafly2011}. From the spectroscopy we have heliocentric redshifts $z_{\mathrm{helio}}$, from which we compute CMB-frame redshifts $z_{\mathrm{CMB}}$ using the angular coordinates of the galaxy and the measured CMB dipole from \cite{Planck2020I}.

We also further cross-match our input data with the \cite{Tempel2011}, \cite{Tempel2014} and \cite{Tempel2017} catalogues to obtain group-averaged CMB-frame redshifts $z_{\mathrm{group}}$ and morphologies $M$, where available.\footnote{In doing so, we recompute the group-averaged redshifts ourselves to avoid the additive redshift approximation used in \cite{Tempel2017}.} The Tempel groups are constructed using SDSS DR12 data down to a limiting $r$-band Petrosian magnitude of 17.77 and the robust Friends-of-Friends algorithm \citep{Turner1976}. Although this is slightly older data that we use to construct the PV sample, it is substantially deeper than our magnitude limit, and the $z$\,$\leq$\,0.1 data in SDSS is almost identical between DR12 and DR14 --- only 418 of our galaxies are not found within the \cite{Tempel2017} group catalogues, for which we assume no group membership.

Given all this input data, as well as uncertainties for these parameters, we are then able to compute measured values and uncertainties for the FP, our distance indicator. The FP relation has the form
\begin{equation}
    \log R_e = a \log \sigma_0 + b \log I_e + c
\label{eq:FP_relation}
\end{equation}
where $R_e$ is the effective radius (in $h^{-1}$\,kpc) and is derived from two quantities, the angular effective radius $\theta_e$ (in arcsec), which can be measured from our photometry, and the distance, which is the desired quantity. The distance-independent quantities in the FP relation are the central velocity dispersion $\sigma_0$ (in km\,s$^{-1}$) and the mean surface brightness within the angular effective radius, $I_e$; these can be measured from spectroscopy and photometry respectively. The coefficients of the FP are represented as $a$, $b$ and $c$. It is common for the FP to be written in shorthand form as $r = as + bi +c$, where $r=\log R_{e}$, $s=\log \sigma_{0}$, and $i=\log I_{e}$, a convention we also adopt in this work.

To start, we convert the de Vaucouleurs scale radius to an angular effective radius in arcseconds $\theta_{e}$, using
\begin{equation}
\theta_{e} = r^{\mathrm{deV}}\sqrt{b/a}.
\label{eq:FPtheta}
\end{equation}
From this, we compute the distance-dependent quantity of the FP, the physical effective radius in units of $h^{-1}\,\mathrm{kpc}$
\begin{equation}
    r_{z} = \log(\theta_{e}) + \log(d(z_{\mathrm{group}})) - \log(1+z_\mathrm{helio}) + \log(\pi/648),
    \label{eq:FPr}
\end{equation} 
where $d(z_\mathrm{group})$ is the comoving distance in $h^{-1}$ Mpc to the group redshift under our assumed fiducial cosmology. We have specifically included the subscript in $r_{z}$ to highlight that this is the physical effective radius inferred using the observed redshifts rather than the true comoving distance to the galaxy. The last term accounts for the conversion from arcseconds to $h^{-1}\,\mathrm{kpc}$, and we have been careful to distinguish between the use of group redshifts for the comoving distance calculation, and heliocentric redshifts for the conversion from comoving to angular diameter distance \citep{Calcino2017}. In using the \cite{Tempel2017} group redshifts for Eq.~\ref{eq:FPr}, we are neglecting intra-group peculiar velocities, which helps reduce non-linearities in the final peculiar velocity measurements.

The second, distance-independent, FP parameter, the effective surface brightness in units of $L_{\odot}\,\mathrm{pc}^{-2}$, is computed from the same angular effective radius, along with the apparent magnitude and a similar unit conversion,
\begin{multline}
    i  = 0.4(M_{\odot}^{r} - m^{\mathrm{deV}}_{r} - 0.85z_{\mathrm{group}} + k_{r} + A_{r}) - \log(2\pi\theta_{e}^{2}) \\
    + 4\log(1+z_{\mathrm{helio}}) + 2\log(64800/\pi).
    \label{eq:FPi}
\end{multline}
The additional $4\log(1+z_{\mathrm{helio}})$ factor accounts for surface brightness dimming, while $0.85z_{\mathrm{group}}$ is an evolution correction following \cite{Bernardi2003b}. $k_{r}$ is the k-correction computed using the heliocentric redshift and $g-r$ colour following \cite{Chilingarian2010}, while $M_{\odot}^{r}=4.65$ is the  absolute magnitude of the Sun in the SDSS $r$-band \citep{Willmer2018}.

The last of our three parameters, the velocity dispersion in $\mathrm{km\,s^{-1}}$ is simply derived as $s \equiv \log(\sigma_{0})$ where $\sigma_{0}$ is the central velocity dispersion. This is obtained from the measured velocity dispersion using the aperture correction of \cite{Jorgensen1995} such that
\begin{equation}
    s = \log(\sigma) - 0.04 (\log(\theta_{e}) - \log(8\theta_{\rm ap})),
    \label{eq:FPs}
\end{equation}
where $\theta_{\rm ap}$ is the fibre radius used to obtain the galaxy spectrum. $\theta_{\rm ap}=1.5 \arcsec$ for objects with SDSS plate number < 3510, whereas $\theta_{\rm ap}=1.0 \arcsec$ for objects observed on later plates, as this demarcates the transition from the older 640-fibre SDSS spectrograph to the newer 1000-fibre BOSS spectrograph.\footnote{As detailed here: \url{https://www.sdss.org/dr12/spectro/spectro_basics/.}}

Eqs.~\ref{eq:FPr}-\ref{eq:FPs} form the basis of our input catalogue and are used to fit the FP and in turn obtain peculiar velocities. An overview of how this is done is covered in the next sub-section, with detailed descriptions given in Sections~\ref{sec:FP} and Sections~\ref{sec:logdist}.

\subsection{From FP parameters to peculiar velocities}
\label{sec:FPtoPV}

The measured parameters $r_{z}$, $s$, and $i$ form the input for the FP. Because the observed group redshift has been used to obtain $r_{z}$, this does not necessarily represent the true intrinsic size of the galaxy, which we denote as $r_{t}$ --- the peculiar velocity of the group in which the galaxy resides creates a difference between $r_{z}$ and $r_{t}$ equal to the log-distance ratio, $r_{z}-r_{t} = \eta \equiv \mathrm{log}(d(z_{\mathrm{group}})/d(\mathrm{\bar{z}}))$. This follows from a comparison of how the physical size in Eq.~\ref{eq:FPr} would be computed from the angular size if one were to use either the (known) group or (unknown, but desired) cosmological redshift \citep{Springob2014}. \footnote{This equation differs slightly from that in \cite{Springob2014}, which includes an additional factor $\mathrm{log}(1+\bar{z})-\mathrm{log}(1+z_{\mathrm{group}})$. However, this factor arises due to a common misconception in the conversion from angular diameter distance (which is proportional to the difference in effective radii) to comoving distance. Although $r$ is an physical size, as pointed out in \cite{Calcino2017} the conversion from comoving distance to luminosity or angular diameter distance always depends on \textit{heliocentric} redshift, even when estimating the comoving distance from a distance indicator. As a result, the additional factor included by \cite{Springob2014} actually vanishes. Fortunately, the effect of incorrectly including this term only produces a relative error in each galaxy's log-distance ratio and peculiar velocity $v_p$ that is approximately equal to two times its redshift: $\Delta v_p / v_p \approx \mathrm{ln}(10) z/(1+z)$ (see Appendix~\ref{sec:incorrect}). For the 6dFGSv this means that the the resulting errors in the measured peculiar velocities or log-distance ratios given by \cite{Springob2014} are generally $<$10\% of the typical statistical uncertainty.}

Hence, the peculiar velocity of an object is derived from its offset from the best-fit FP in the $r$ direction --- the FP is fit using the ensemble of measured $r_{z}$, $s$, and $i$ and for each galaxy can be used to predict $r_{t}$ and then $\eta$. Using $\eta$ is convenient because if the FP is treated as Gaussian, and the uncertainties in each of the input parameters are Gaussian, then $\eta$ is also Gaussian distributed (modulo some small skewness introduced due to selection functions and other complexities in the fitting process). This is the main quantity provided in the SDSS PV catalogue.

From $\eta$ we can compute the distance modulus,
\begin{equation}
    \mu = 5\mathrm{log}(d(z_{\mathrm{CMB}})) - 5\eta + 25,
\end{equation}
where $d(z_{\mathrm{CMB}})$ is taken to be in units of Mpc. The peculiar velocity can be computed fully by first numerically inverting the redshift-distance relation to convert from log-distance ratio to cosmological redshift $\bar{z}$, then using the equation for propagating redshifts to obtain the peculiar velocity \citep{DavisT2014}. The downside of this procedure is that the distribution for the peculiar velocity is no longer Gaussian and so must be expressed as a non-Gaussian PDF (taking into account the Jacobian of the transformation; \citealt{Johnson2014}), approximated (i.e., \citealt{Scrimgeour2016}) or corrected for in some other manner (e.g., \citealt{Hoffman2021, Qin2021}).

An alternative, which is used in this work whenever a peculiar velocity is presented, is to use the approximate conversion \citep{Watkins2015}
\begin{equation}
    v_{p} \approx \frac{cz_{\mathrm{mod}}}{1+z_{\mathrm{mod}}}\mathrm{ln}(10)\eta,
\end{equation}
where
\begin{equation}
    z_{\mathrm{mod}}=z_{\mathrm{CMB}}[1+1/2(1-q_{0})z_{\mathrm{CMB}}-1/6(j_{0}-q_{0}-3q_{0}^{2}+1)z^{2}_{\mathrm{CMB}}],
\end{equation}
$c$ is the speed of light, and $q_{0}$ and $j_{0}$ are the present day deceleration and jerk parameters, which for our fiducial cosmology have the values $q_{0}=-0.535$ and $j_{0}=1$. This estimator retains the same distribution as the original log-distance ratio. It is derived from a Taylor expansion of $\mathrm{ln}(1-v_{p}/cz_{\mathrm{CMB}})$ and so is accurate as long as the true peculiar velocity (not necessarily the measured value) satisfies $v_{p} \ll cz_{\mathrm{CMB}}$. Given the low-redshift cut applied to the SDSS PV data ($z_{\mathrm{CMB}}>0.0033$), we expect this approximation to work well---\cite{Howlett2017c} demonstrated that this estimator holds even for the much lower redshift 2MASS Tully-Fisher catalogue.

\subsection{Contamination by spirals and other interlopers}
\label{sec:spirals}

Before fitting the FP, there are a few caveats and additional steps that must explored. Our selection of 42,170 galaxies in Section~\ref{sec:selection} is designed to select photometrically-clean red E/S0 galaxies with no H$\alpha$ emission. However, this is not sufficient to isolate a clean sample for the FP analysis. Additional steps are needed to remove unsuitable objects that can act as outliers from the FP, resulting in spurious velocity measurements for these objects and a potential bias in the fit to the FP itself. 

In order to identify and remove these interlopers, we used the morphological classifications from GalaxyZoo \citep{Willett2013} and 
\cite{Tempel2011,Tempel2014}, and we visually inspected all galaxies on 
1\,$\times$\,1 arcmin colour cutouts extracted from the Pan-STARRS1 
\citep{Chambers2016} and Legacy Survey images \citep{Dey2019}. The 
deeper images of the Legacy Survey, particularly the model residual 
images, allow considerably better discrimination than was possible with 
previous survey images. We identified the following unsuitable objects:
\begin{enumerate}
    \item{spiral galaxies that had not been previously recognised in the shallower images;}
    \item{galaxies where the measurement of the FP photometric  parameters is likely to be unreliable due to overlapping sources (either stars or other galaxies);}
    \item{galaxies with strong asymmetries;}
    \item{galaxies with strong central dust features which are likely to bias the velocity dispersion measurements.}
\end{enumerate}
While all visual inspection work is subjective, our aim was to identify galaxies that are clearly not suitable to be included in our FP analysis.

\begin{figure}
\centering
\includegraphics[width=0.48\textwidth, trim=0pt 0pt 0pt 0pt, clip]{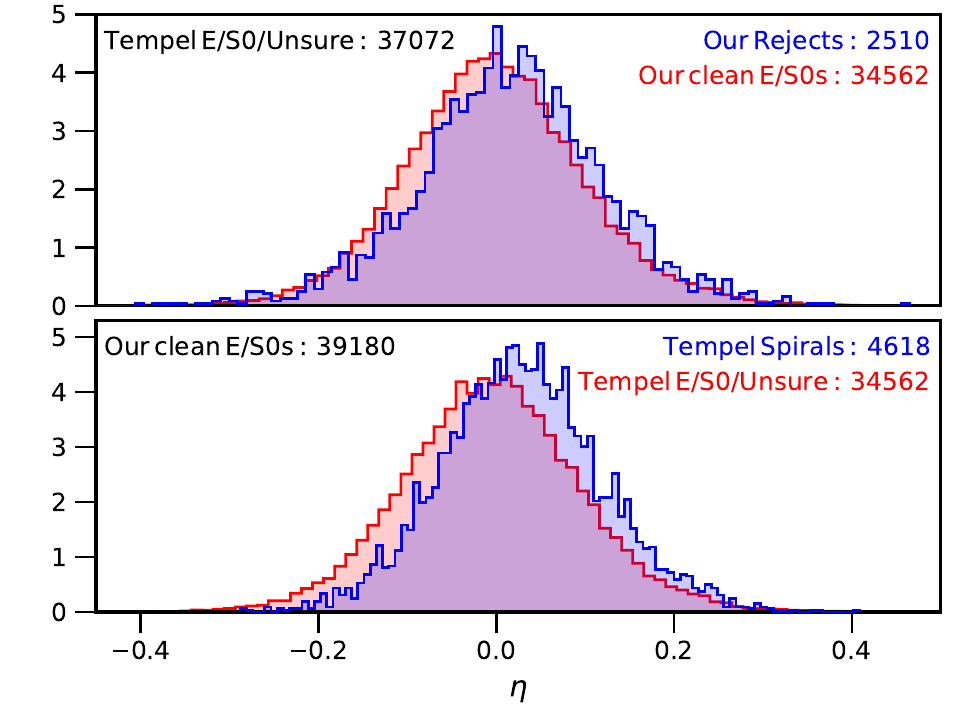}
  \caption{Normalised histograms of the best-fit log-distance ratios for the SDSS PV sample using different E/S0 classifications. The top panel shows measurements for all galaxies classified by \protect\cite{Tempel2014} as early-type/unsure, and further subdivided into those we keep (red) or reject (blue) after our visual inspection. The bottom panel shows the opposite; galaxies classified by us as photometrically clean early-types split into their \protect\cite{Tempel2014} classification. In both cases there is a remaining subset of interlopers with log-distance ratios offset from the global mean, requiring us to adopt an either/or classification to remove. Note that the red histograms in both panels, despite containing the same numbers of objects, are not quite the same because the FP and log-distance ratios are fit to the full sample of galaxies (red + blue histograms), which differ between the two panels.}
  \label{fig:spirals}
\end{figure}

Considering first the binary classifications provided by \cite{Tempel2014}, we remove 5,098 objects confidently identified as spirals ($M$=1), fit the FP and derive log-distance ratios (following the procedures detailed in Sections~\ref{sec:FP} and~\ref{sec:logdist}). We then compare histograms of the log-distance ratios for the remaining objects based on whether they are flagged as E/S0s or rejected by our visual identification. As shown in the top panel of Figure~\ref{fig:spirals}, we find that a remaining 2,510 galaxies visually rejected by us pass the \cite{Tempel2014} criteria and exhibit log-distance ratios strongly offset from our clean E/S0 classification. These would hence also need to be removed. 

We then play the reverse game, removing the 2,990 galaxies that \textit{we} have visually rejected before fitting the FP and deriving log-distance ratios. Unfortunately, as shown in the bottom panel of Figure~\ref{fig:spirals}, we then find a remaining 4,618 objects that we classify as clean E/S0s but \cite{Tempel2014} classify as spirals that also exhibit biased log-distance ratios. From these results it is clear that both our classification and the \cite{Tempel2014} classification are picking up different subsets of interlopers, both of which bias our sample. This can be verified by looking at the FP parameters themselves (in Table~\ref{tab:FP}, see also Fig~\ref{fig:SDSS_FP}); our rejects are situated primarily at the large radius, low surface brightness end of the FP, whilst \cite{Tempel2014} spirals are brighter and more compact. As such, we adopt a hybrid classification, removing a combined total of 7,608 galaxies from our sample if \textit{either} we or \cite{Tempel2014} have classified the galaxy as a spiral/reject. It is worth noting that we did test using cuts in the GalaxyZoo probability itself for cleaning the sample, but ultimately found that the combination of binary classifications removed the contamination from spirals whilst retaining a larger total number of objects. The number of objects remaining after removing spirals/rejects is 34,562.

\subsection{Angular Mask}
\label{sec:mask}

The complete SDSS PV catalogue contains galaxies spread over the contiguous area of the SDSS Northern Galactic Cap. However, the data comes from a combination of various SDSS releases up to and including Data Release~14 (DR14). In order to produce random and mock galaxy catalogues that reproduce the angular distribution of the data, we require an angular mask that describes which regions of sky we expect to be present or missing from the catalogue due to how the data was collected. However, the SDSS PV data is a complex combination of data from the SDSS DR8, plus additional low redshift objects from later data releases. To the best of our knowledge, there are currently no publicly available angular masks describing the distribution of DR8 galaxies across the SDSS footprint that take into account holes arising from centreposts, bright stars, or the tiling geometry of the survey, let alone when newer data up to DR14 is included. All of these are relevant for samples where the clustering may be measured.

\begin{figure*}
\centering
\includegraphics[width=\textwidth, trim=0pt 0pt 0pt 0pt, clip]{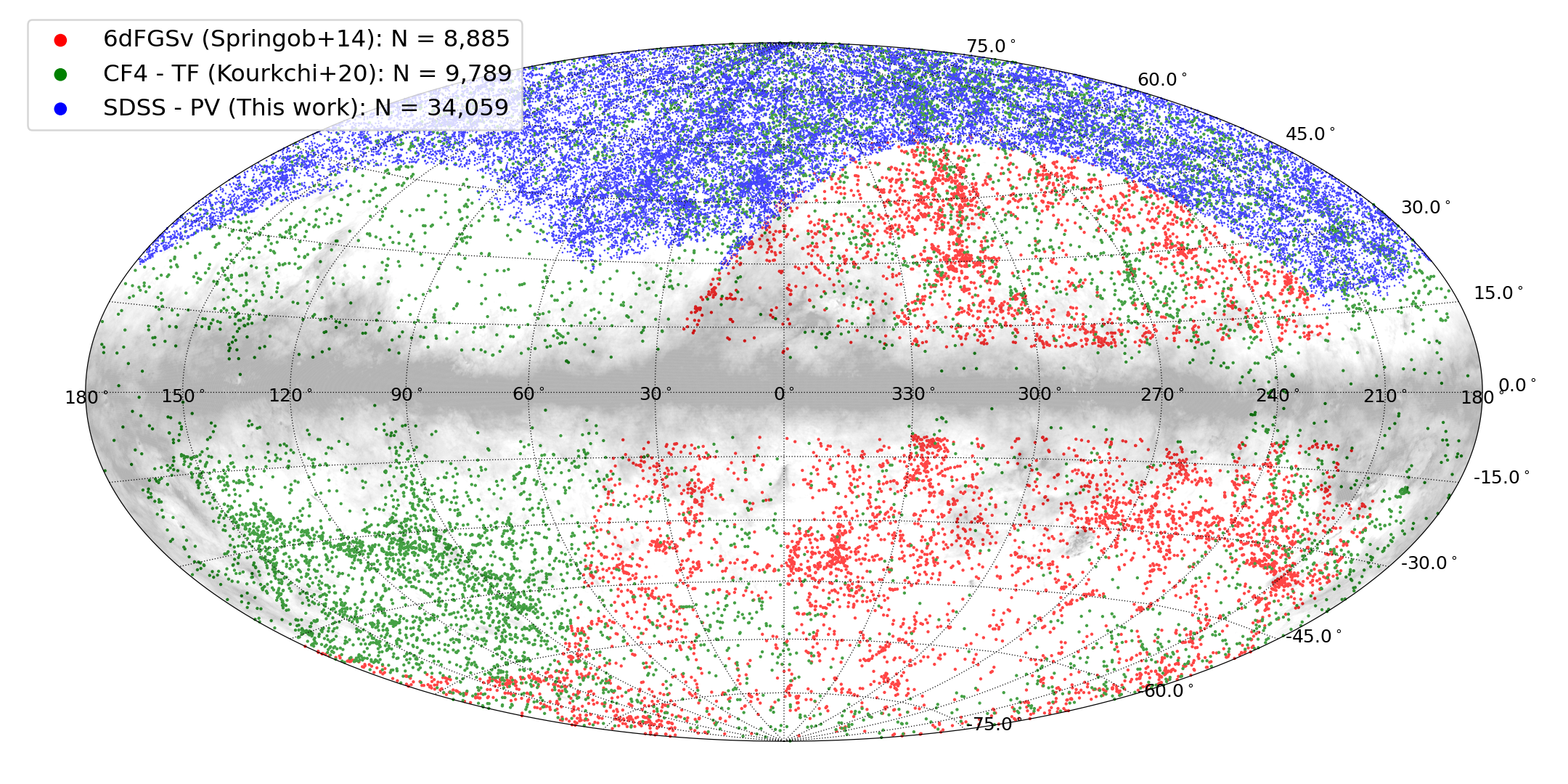}
  \caption{The distribution of the SDSS PV data (blue) in Galactic coordinates compared to data from the 6-degree Field Galaxy Survey (red; \citealt{Springob2014}) and Cosmicflows-4 TF (green; \citealt{Kourkchi2020}). The shaded region shows areas of high galactic extinction around the plane of the Milky Way, grey-scaled by $\log (E(B-V))$.}
  \label{fig:SDSS_angular}
\end{figure*}

As such, rather than tackling the difficult task of tracking back the imaging, targeting, tiling and veto masks for the SDSS PV catalogue, we instead opt for the simpler solution of identifying the SDSS PV galaxies that belong to a pre-existing, well-defined, angular mask and using only those galaxies for scenarios where the angular distribution may be important (for example in measuring the clustering of the galaxy density field). For this purpose, we use modified versions of the \textsc{mangle} \citep{Hamilton2004, Swanson2008} polygon files provided alongside the NYU Value-Added Galaxy Catalogue Data Release~7 (DR7;  \citealt{Blanton2005}).\footnote{Available here: \url{http://sdss.physics.nyu.edu/vagc/}} We then identify and flag any galaxies that are outside this mask, and exclude them from clustering measurements and any computation requiring knowledge of the precise sky coverage of the data (such as computing the number density per unit volume). It should be noted that all galaxies are still retained in the SDSS PV catalogue and have measured peculiar velocities; those that are within the mask are merely flagged in the data file as \textsc{in\_mask}. Of the $34,562$ early-type galaxies retained in the sample so far, $944$ are outside the DR7 footprint; a small fraction that we expect to have little bearing on further analysis of the data. The total number of SDSS PV catalogue galaxies in the mask is thus $33,618$ and the total area covered by the angular mask is 7016\,deg$^2$; the sky coverage of the SDSS PV catalogue relative to existing peculiar velocity data from 6dFGSv and the CF4-TF sample is shown in Galactic coordinates in Figure~\ref{fig:SDSS_angular}.

\subsection{Random catalogue}
\label{sec:randoms}

\begin{figure}
\centering
\includegraphics[width=0.48\textwidth, trim=0pt 0pt 0pt 0pt, clip]{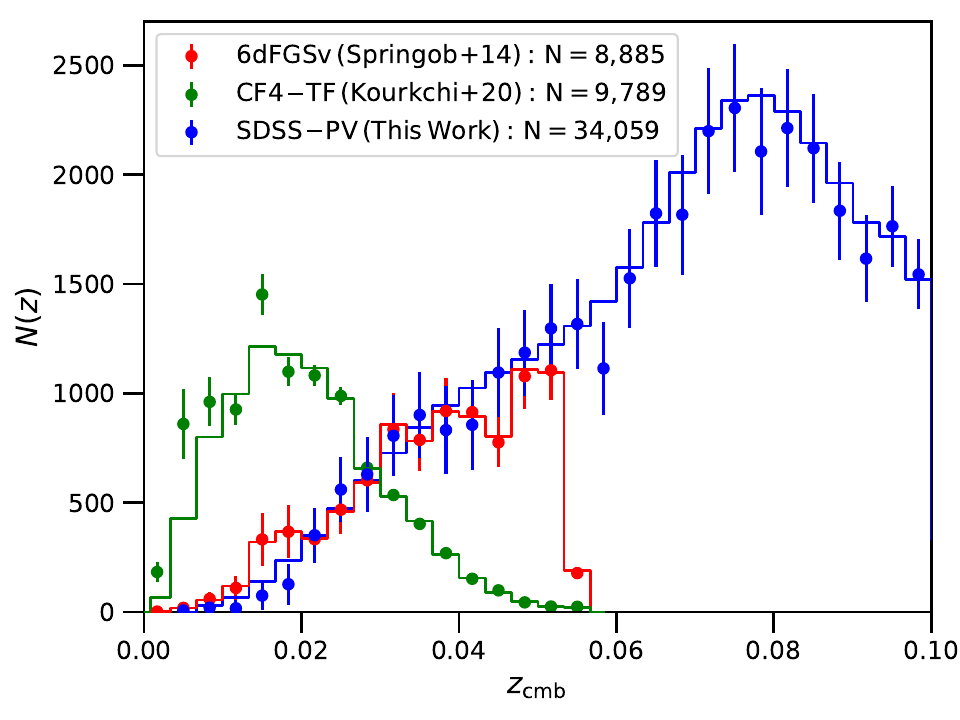}
  \caption{The redshift distribution of the SDSS PV sample compared to the 6-degree Field Galaxy Survey (red; \protect\citealt{Springob2014}) and Cosmicflows-4 TF (green; \protect\citealt{Kourkchi2020}). Points show the number of galaxies per bin of width 1000\,km\,s$^{-1}$, error bars encapsulate the cosmic variance from ensembles of simulated mock surveys (see Section~\ref{sec:mocks} for our SDSS PV mocks and \protect\citealt{Qin2019b} and \protect\citealt{Qin2021b} for 6dFGSv and CF4-TF mocks respectively), and lines are the mean distributions from these mocks.}
  \label{fig:SDSS_nz}
\end{figure}

In order to measure the clustering of the SDSS PV sample, both in this and in future work, we require a random unclustered sample of data points that can be used to compute the relative overdensity of the galaxies. Given the angular mask, we first use \textsc{mangle} to produce a random sample of points that match the angular distribution of the data. The redshift distribution of the data was then fit with a smoothed spline to capture the general shape of the survey selection function without incorporating real large-scale structure along the line-of-sight. We then used this smoothed spline to sample redshifts for the random catalogues, and to down-sample the simulated galaxy catalogues presented in Section~\ref{sec:mocks}. The number of SDSS PV galaxies as a function of redshift is shown in Figure~\ref{fig:SDSS_nz} alongside the 6dFGSv and CF4-TF data. We can see that the three samples are highly complementary, with the CF4-TF sample peaking at lower redshift than the 6dFGSv or SDSS PV samples. The two FP samples contain a similar number of objects at $z<0.055$ (although not the same number density; 6dFGSv is spread over more than twice the area of the SDSS PV), but the SDSS data extends to higher redshift, up to $z<0.1$, where the bulk of our new peculiar velocity measurements lie. This figure emphasises that our new data push into a previously unexplored redshift regime, although it is certainly important to point out that the improvements in terms of constraining power from SDSS compared to previous datasets are more modest than the number of galaxies would suggest, as the points at higher redshifts have proportionately larger errors on their peculiar velocities (see Sections~\ref{sec:logdist} and ~\ref{sec:cosmo}).

\section{Mock Galaxy Surveys}
\label{sec:mocks}

Mock galaxy catalogues (mocks) that reproduce the clustering and selection function of the peculiar velocity data are essential for validating the methodology used to extract peculiar velocities from a distance indicator in the presence of survey selection effects. They are also a necessity for the interpretation of cosmological results from the data. In this section, we describe the production of 2,048 mocks for the SDSS PV data that are designed to reproduce all relevant aspects of the data while encapsulating the effects of cosmic variance on the peculiar velocity field.

\subsection{Characterising the SDSS data}
\label{sec:character}

To begin, we first derive a series of simple fitting functions for the properties of the SDSS data that will enable us to add realistic uncertainties and observed quantities to the simulations.

\subsubsection{Measurement uncertainties}

We start by looking at the errors in the measured FP parameters. We denote the errors on $r_{z}$, $s$ and $i$ as $e_{r}$, $e_{s}$ and $e_{i}$ respectively. They are derived from the errors on $r^{\mathrm{deV}}$, $b/a$, $m_{r}^{\mathrm{deV}}$ and $\sigma$ provided with the publicly available SDSS data by error propagation through Eqs.~\ref{eq:FPtheta}-\ref{eq:FPs}. 

First, as the fainter galaxies in our sample have larger uncertainties in their apparent magnitude, we find (via Eq.~\ref{eq:FPi}) a strong correlation of the surface brightness uncertainties ($e_{i}$) with apparent magnitude. The distribution of $e_{i}$ is found to be close to log-normal, so we instead quantify the relationship between de~Vaucouleurs $r$-band magnitude and the logarithm of the uncertainty in the surface brightness, $\log(e_{i})$, by binning the data in magnitude bins and fitting the mean and standard deviation in each bin. $e_{i}$, being the measurement error, can only be positive and hence can be treated in this manner. So although the form of this quantity may appear strange ($\log(e_{i})$ being the log of the error in a quantity that is already logarithmic), we use this purely as it is more Gaussian distributed than the uncertainty itself, and hence allows us to randomly assign errors to the mocks more easily (fitting the $e_{i}$ as a function of apparent magnitude instead and then drawing from a log-normal distribution would be mathematically equivalent).

\begin{figure*}
\centering
\includegraphics[width=0.33\textwidth, trim=0pt 0pt 0pt 0pt, clip]{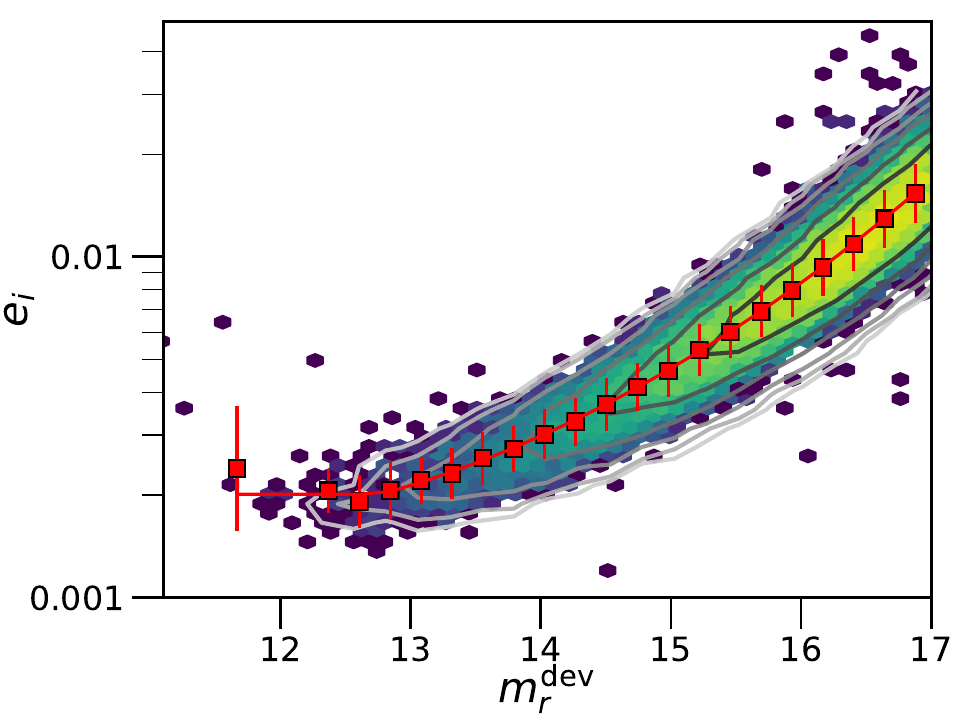}
\includegraphics[width=0.33\textwidth, trim=0pt 0pt 0pt 0pt, clip]{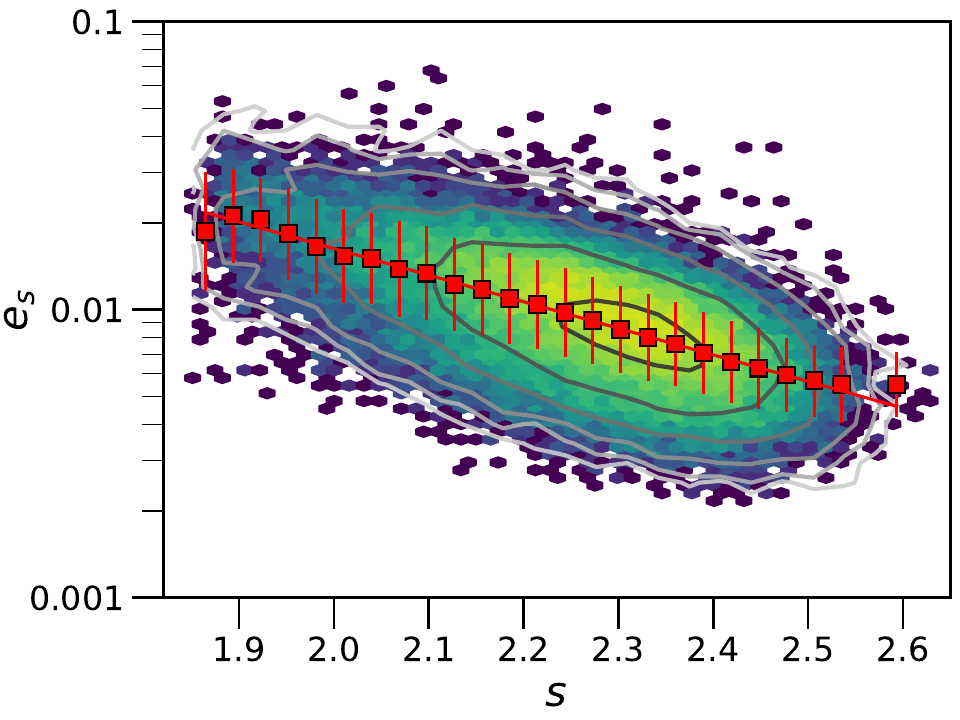}
\includegraphics[width=0.33\textwidth, trim=0pt 0pt 0pt 0pt, clip]{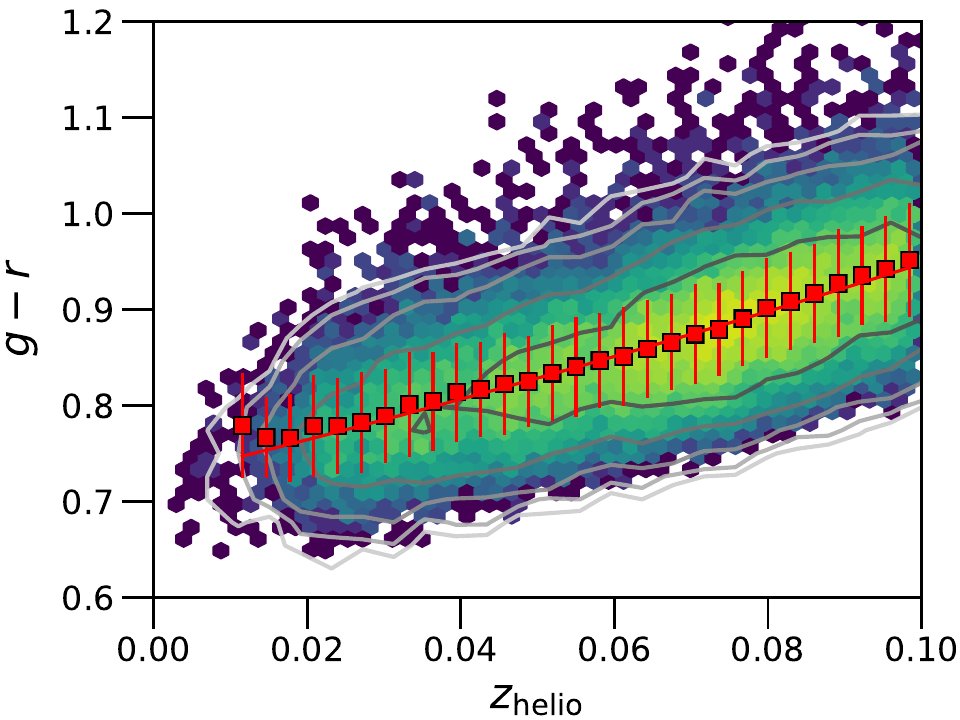}
  \caption{Characterisation of the measurement errors in surface brightness and velocity dispersion ($e_{i}$ and $e_{s}$ respectively) and $g-r$ colours in the SDSS PV sample. In each panel the real data are shown as a hex-binned distribution colour-coded by the number of galaxies in each bin. The contours show a random distribution of points drawn using the $x$-axis values of the real data (de~Vaucoleurs $r$-band magnitude, velocity dispersion and heliocentric redshift respectively from left to right) and the fitting formulae provided in Section~\ref{sec:character}. The points in each panel show the mean and scatter to which these formulae were fit and the red lines show the fit itself.}
  \label{fig:SDSS_character}
\end{figure*}

We find that the mean is well represented by a piece-wise function that is quadratic for faint galaxies and asymptotes to a constant for galaxies brighter than some limiting magnitude. The Gaussian scatter, denoted $\sigma_{\log(e_{i})}$ is fit well by a straight line. Our best-fit relationship is given by
\begin{align}
    \langle \log(e_{i}) \rangle &= 
    \begin{cases}
        -2.35 & m_{r}^\mathrm{deV} < 12.77 \\
        0.02(m_{r}^\mathrm{deV})^{2} - 0.43m_{r}^\mathrm{deV} - 0.13 & \mathrm{otherwise}
    \end{cases}\\
    \sigma_{\log(e_{i})} &= 0.0034m_{r}^\mathrm{deV}+0.0052 ~.
\end{align}
The data, binned mean and scatter, best-fit relationship and an example set of random values generated using the above fitting formulae are shown in the left panel of Figure~\ref{fig:SDSS_character}.

By construction, \cite{Said2020} set the uncertainty on the effective radius $r$ to be equal to half the uncertainty on the surface brightness. They also set the correlation coefficient between $r$ and $i$ to be $-$1.0, i.e.\ perfectly anti-correlated given they are produced using the same data, but have the opposite dependence on angular effective radius. We adopt the same procedure here, so $e_{r}=0.5e_{i}$. 

The only remaining measurement uncertainty to quantify is the error on the velocity dispersion $e_{s}$. Comparing to the properties of the input data, we find that this is correlated strongly with spectral S/N---unsurprisingly, velocity dispersions can be measured more accurately in spectra with high S/N ratio. This in turn introduces a correlation between velocity dispersion and apparent magnitude because brighter objects can reach a higher spectral S/N in fixed observing time. However, for the purposes of generating the simulations, we do not have spectral S/N ratios from which we could draw velocity dispersion uncertainties. Furthermore, we opt \textit{not} to use apparent magnitude for this, as to do so would introduce correlations between the effective radius/surface brightness and velocity dispersion that are observational rather than intrinsic in nature. In fitting the FP in Section~\ref{sec:FP} we treat the correlation between photometric and spectroscopic observational errors as zero, and using apparent magnitude to generate mock velocity dispersion errors would run counter to this. 

We instead make use of a third correlation found between the error on the velocity dispersion and the velocity dispersion itself, in the same way as was done in \cite{Scrimgeour2016} and \cite{Qin2018}. This correlation arises in combination with those previously identified -- larger velocity dispersions can be measured more precisely both because it is easier to fit the absorption features, but also because the galaxies are, by virtue of the FP, brighter and have higher spectral signal-to-noise. This choice has the benefit of relying only on a spectral property, and one which we produce as part of our simulations.

To produce a fit, we follow the same method as before. We again verify that $e_{s}$ is close to log-normally distributed, and so work with $\log(e_{s})$ and find a good fit using the relation
\begin{align}
    \langle \log(e_{s}) \rangle &= -0.766s - 0.121 \\
    \sigma_{\log(e_{s})} &= -0.049s + 0.236 ~.
\end{align}
The data, fitted relationship and an example distribution drawn from this fit are shown in the middle panel of Figure~\ref{fig:SDSS_character}.

\subsubsection{Colours and $k$-corrections}

The final aspect of the data we need to characterise are the $k$-corrections. We aim to reproduce these in our simulations so that we can be sure that these are not impacting our recovered peculiar velocities for galaxies close to the edge of the magnitude limit, where a small change in the $k$-correction can potentially scatter a galaxy in or out of our sample.

$k$-corrections in the SDSS PV data are obtained using the bivariate fits from \cite{Chilingarian2010} as a function of $g-r$ colour and heliocentric redshift. To replicate these in simulations, we need to assign $g-r$ colours to our mock galaxies based on some other properties that we already have to hand. Looking at the correlations between various aspects of the data, we find that the colour is strongly correlated with redshift. Such a trend can be inferred from Fig.~\ref{fig:colourmag} due to the way intrinsically fainter, bluer galaxies fall out of the sample as we go to higher redshifts. We also found a weaker correlation with velocity dispersion, but that was difficult to model well due to large scatter, and not clearly causative because the velocity dispersion itself also increases with redshift. For simplicity in producing the mocks, we therefore just fit the colour as a function of heliocentric redshift, using a log-normal distribution as above. We note that this has the benefit of simply reducing the $k$-corrections used in the mocks to a univariate redshift-dependent function. Our best-fit relationship is given by
\begin{align}
    \langle \log(g-r) \rangle &= 1.167z_{\mathrm{helio}} - 0.140 \\
    \sigma_{\log(g-r)} &= -0.042z_{\mathrm{helio}} + 0.028  ~.
\end{align}
The right-hand panel of Figure~\ref{fig:SDSS_character} shows the colours of the true SDSS PV data as a function of the colour modelled with this fit. The data are scattered normally about the expected one-to-one line and well reproduced by a random set of points drawn with the scatter given above. Given we can use these formulae to compute colours, we can then combine these with the same redshifts to calculate a $k$-correction.

Overall, we have been able to effectively reproduce the most important aspects of the SDSS PV data using a small number of fitting formulae. In the next section these fitting formula will be used to create a large ensemble of simulations that fully reproduce the data.

\subsection{Producing the simulations}

Our method for producing the simulations is presented in \cite{Qin2019b}. First, an ensemble of 256 dark matter simulations is produced using the approximate N-body code \textsc{l-picola}. Each simulation consists of $2560^{3}$ dark matter particles in a box of edge length $1800h^{-1}$\,Mpc evolved to $z=0$. Dark matter halos are then identified in the simulation using the 3D friends-of-friends algorithm \citep{DavisM1985} with a minimum of 20 dark matter particles. This corresponds to a minimum halo mass of $\sim 5\times 10^{11}h^{-1}M_{\odot}$. From our simulations, we have catalogues of halo positions, velocities and masses.

Galaxies are placed within these halos using a variant of sub-halo abundance matching \citep{Conroy2006}. However, the approximate nature of the \textsc{l-picola} simulations means that only halos, not sub-halos, can be reliably identified from the dark matter field using the friends-of-friends algorithm. Instead, we artificially add in sub-halos by using a power law with two free parameters ($A$ and $\alpha$) to describe the number of sub-halos, $N_{\mathrm{sub}}$, as a function of the mass ratio between parent halos and sub-halos, $f_{M}$,
\begin{equation}
N_{\mathrm{sub}}(f_{M}) = Af_{M}^{-\alpha} ~.
\end{equation}
Integrating this function between some minimum value of $f_{M}$ and 1 (where we set the minimum using a sub-halo mass equivalent to 20 dark matter particles) gives us the \textit{expected} number of sub-halos in each parent halo. 

However, to account for the observed scatter in the sub-halo to halo mass distribution \citep{Giocoli2008, Elahi2018}, in practice we draw an actual number of sub-halo masses using a Poisson distribution with mean $N_{\mathrm{sub}}$. The sub-halos are then placed within their parent halos using a orbital radius and velocity drawn from the NFW profile \citep{Navarro1997}, using the algorithms/equations in \cite{Cole1996, Robotham2018} and mass-concentration relation from \cite{Prada2012}.

Finally, galaxies are drawn from the FP distribution with the best-fit SDSS parameters given in Table~\ref{tab:FP}. Given our set of halo and sub-halo positions, velocities and masses, these are assigned to the halos and sub-halos based on rank-ordering the masses and the value of $2r+i$ (which acts as the proxy for luminosity). We add scatter to this rank-ordering based on a third free parameter, $\sigma_{\log L}$, between the log of the masses in units of $M_{\odot}$, and the log of the luminosities in units of $L_{\odot}$. Galaxies are given the position and velocity of their host halo/sub-halo.

\subsubsection{Incorporating the selection function}

After FP parameters have been generated, we apply the selection effects and incorporate measurement uncertainties into the mocks. We place 8 separate observers in each of our 256 simulations spaced maximally far apart ($\sim$600$h^{-1}$\,Mpc) and for each one:
\begin{itemize}\vspace*{-6pt}
    \item{Use \textsc{mangle} and the angular mask from Section~\ref{sec:mask} to down-sample the mock catalogues to match the footprint of the data.}
    \item{Perturb the effective radius $r$ for each galaxy based on its peculiar velocity (via the log-distance ratio).}
    \item{Generate $g-r$ colours for the mock galaxies based on the fit to the heliocentric redshift and velocity dispersion $s$ identified in Section~\ref{sec:character}.}
    \item{Generate $r$-band $k$-corrections for each galaxy using the $g-r$ colour, heliocentric redshift and fitting formulae of \cite{Chilingarian2010}}.
    \item{Compute the $r$-band Milky-Way extinction for each galaxy based on its location on the sky using the implementation of the \cite{Schlafly2011} dust maps in the Python \textsc{dustmaps} package \citep{Green2018}.\footnote{https://dustmaps.readthedocs.io/en/latest/}}
    \item{Combine the FP parameters with the $k$-correction, Galactic extinction and luminosity distance to each galaxy to compute the $r$-band apparent magnitude by inverting Eqs~\ref{eq:FPr} and~\ref{eq:FPi}.}
    \item{Use the magnitude and FP parameters to produce correlated uncertainties on $r$, $s$ and $i$ based on the fitting formulae in Section~\ref{sec:character} and then use these to generate \textit{observed} FP measurements centred on the true values.}
    \item{Apply cuts to the simulated observed redshifts, velocity dispersions and magnitudes, matching those in the SDSS data: $0.0033 < z < 0.1$, $s > \log(70)$ and $m_{r}^{\mathrm{deV}}<17.0$.}
    \item{Subsample the observed redshifts of the mock galaxies using the smooth spline fit to the data from Section~\ref{sec:randoms}. After populating the simulation with galaxies and applying the other selection effects, the number density of galaxies in the catalogues is larger than in the data by an approximately constant factor---the trend as a function of redshift matches the data well due to the inclusion of the magnitude and velocity dispersion cuts, but our model is quite simplistic, is fit only to the monopole of the clustering (see Section~\ref{sec:tuning}), and does not account for things like the redshift success rate. All of these effects cause the number density obtained from the previous steps to be larger than in the data, which we remedy by randomly subsampling. Doing this randomly ensures the clustering properties of the mocks remain unchanged by the subsampling.}
\end{itemize}

\subsubsection{Tuning the mocks}
\label{sec:tuning}

Given that the relationships between galaxy properties and uncertainties have been characterised in Section~\ref{sec:character}, our selection function is well defined, and parent halo concentrations are computed based on fits in the literature to high resolution N-body simulations, our entire procedure contains three remaining free parameters, $A$, $\alpha$ and $\sigma_{\log L}$. These are the slope and normalisation of the sub-halo mass ratio distribution, and the scatter between halo/sub-halo mass and galaxy luminosity.

\begin{figure}
\centering
\includegraphics[width=0.48\textwidth, trim=0pt 0pt 0pt 0pt, clip]{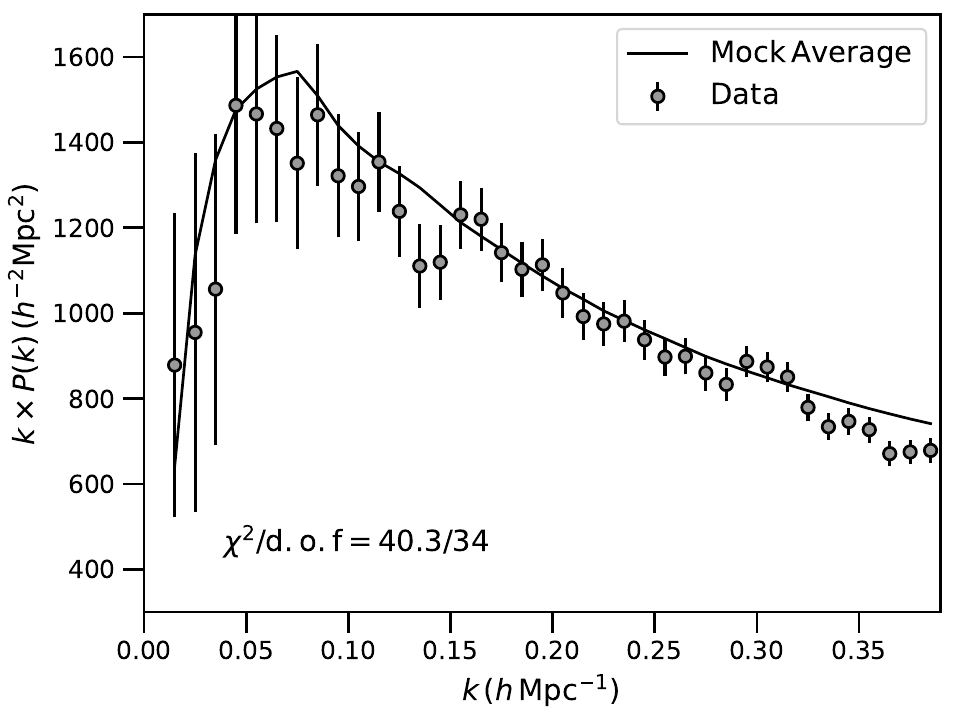}
  \caption{The spherically-averaged density power spectrum of the SDSS PV sample (points) compared to the mean of the mocks (line). The error bars come from the variance in the mocks, which captures both cosmic variance and shot noise. The chi-squared value is calculated from the difference between the data and the mock-mean, and shows that the mocks reproduce the clustering in the data well.}
  \label{fig:pk}
\end{figure}

These parameters are tuned by fitting the monopole of the SDSS PV galaxy density power spectrum. We optimise by brute-force; iteratively populating the halo catalogues, applying the selection function, then computing the galaxy power spectrum and computing the $\chi^{2}$ relative to the SDSS PV power spectrum. As the covariance matrix for this comparison itself has to be computed from mocks, we repeat this entire optimisation process iteratively; using a first guess for the free parameters to generate the first set of mocks and covariance matrix, which is then subsequently used to tune the next generation of mocks. We perform the entire procedure over 4 generations, after which the best-fit parameters do not change considerably between successive generations. In the end, we find best-fit values of $A=1.850$ and $\alpha=1.175$ for the power law amplitude and slope respectively, and $\sigma_{\log L}=0.138$ for the scatter. These are comparable to the values found for 6dFGSv \citep{Qin2019b}, but with lower slope and higher amplitude. This is consistent with our finding that the clustering amplitude (which is mainly set by $\alpha$) is lower for SDSS than 6dFGSv, but the number density (set mainly by the overall normalisation, $A$) is higher. After iterating, the clustering of the mocks reproduces the data well, as shown in Figure~\ref{fig:pk}. The $\chi^{2}$ difference between the power spectrum of the data and the average of the mocks, computed using the covariance matrix estimated from the mocks, is $40.3$ for $34$ degrees of freedom.

In the following section, we explain how we fit the FP from the data and the mocks, and demonstrate that the mocks satisfactorily reproduce the expected distribution of the SDSS data.

\section{Fitting the Fundamental Plane}
\label{sec:FP}

We calculate log-distance ratios and peculiar velocities from the SDSS PV FP data using the Maximum Likelihood Gaussian method introduced in \cite{Saglia2001} and \cite{Colless2001} for analysis of the EFAR sample, and as also used for 6dFGSv data \citep{Magoulas2012,Springob2014}. This is a three-step process, where first the FP itself is fit to the data assuming no peculiar velocities, then the offset from the best-fitting plane is used to infer the peculiar velocity of the individual galaxies. Finally, the zero-point of the FP is calibrated, which is akin to correcting for the fact that the first step of the process assumed an average bulk motion of zero across the \textit{entire} SDSS PV sample, which is unlikely to be true in reality. We note that the procedure can actually all be carried out in a single Bayesian hierarchical model, as done in \cite{Dam2020} and \cite{Said2020}. However this is computationally expensive and so works best when the goal is to evaluate the posterior for a smaller number of derived parameters (such as the growth rate of structure or bulk flow), rather than to produce a catalogue of individual velocities for each galaxy. Nonetheless, this is a clear place for future work to improve on.

For the first two stages we assume the FP is described by a censored 3D Gaussian, so that the probability of observing $N$ galaxies with effective radii, velocity dispersions and surface brightnesses $\boldsymbol{x}_{n} = \{r_{n}$, $s_{n}$, $i_{n}\}$ can be written
\begin{equation}
\mathcal{L} = \prod_{n=1}^{N}\biggl(\frac{1}{(2\pi)^{3/2}|\boldsymbol{\mathsf{C}}_{n}|^{1/2}f_{n}}\exp\biggl[-\frac{1}{2}(\boldsymbol{x}_{n}-\boldsymbol{\bar{x}})\boldsymbol{\mathsf{C}}_{n}^{-1}(\boldsymbol{x}_{n}-\boldsymbol{\bar{x}})^{T}\biggl]\biggl)^{1/S_{n}}
\label{eq:like}
\end{equation}
where $\boldsymbol{\bar{x}} = \{\bar{r}$, $\bar{s}$, $\bar{\imath}\}$ is the mean of the FP and $f_{n}$ normalises the likelihood of the observed galaxy $n$ to 1 over the observed parameter space; $f_{n}<1$ unless the data is uncensored and contains no selection effects. $S_{n}$ is an inverse weighting to account for galaxies that are missing from our sample due to the selection function and is based on the commonly used $1/V_{\mathrm{max}}$ weighting \citep{Schmidt1968}. The full likelihood would properly require calculation of the integral of the PDF times the selection probability for each galaxy and FP model. This is computationally expensive, and one can instead use a $1/S_{n}$ weighting to approximate the full likelihood. As discussed in \cite{Eadie1971}, this results in the same maximum likelihood FP values, but underestimates the variance in the parameters. This is fine for the purposes of our calculation, as we are only interested in the best-fit values for the FP, and estimate the uncertainties using the variance across our ensemble of mock catalogues, which also incorporate the effects of cosmic variance.   

Finally, the covariance matrix $\boldsymbol{\mathsf{C}}_{n}$ describes the scatter in the FP, which consists of both intrinsic scatter $\boldsymbol{\Sigma}$ and measurement uncertainty $\boldsymbol{\mathsf{E}}_{n}$. Assuming both of these individual components are Gaussian and the measurements are unbiased (so that the inclusion of measurement noise does not bias the measurements away from the mean $\bar{x}$), we write this as 
\begin{equation}
    \boldsymbol{\mathsf{C}}_{n} = \boldsymbol{\Sigma} +\boldsymbol{\mathsf{E}}_{n} = 
    \begin{pmatrix}
    \sigma^{2}_{r} & \sigma_{rs} & \sigma_{ri} \\
    \sigma_{rs} & \sigma^{2}_{s} & \sigma_{si} \\
    \sigma_{ri} & \sigma_{si} & \sigma^{2}_{i} 
    \end{pmatrix} + 
    \begin{pmatrix}
    e^{2}_{r}+\epsilon^{2}_{r} & 0 & -e_{r}e_{i} \\
    0 & e^{2}_{s} & 0 \\
    -e_{r}e_{i} & 0 & e^{2}_{i}
    \end{pmatrix} ~.
    \label{eq:covariance}
\end{equation}
The components of the error matrix $\boldsymbol{\mathsf{E}}_{n}$ are obtained directly from the $n^{\mathrm{th}}$ galaxy's measurement uncertainties. We also assume a perfect anti-correlation between $r$ and $i$ (cf.\ Section~\ref{sec:character} and \citealt{Said2020}), no correlation between the photometric and spectroscopically obtained measurements, and add a minimal contribution to the uncertainty on $r$, $\epsilon_{r}=\mathrm{log}(1+300/cz_{n})$, to account for non-linear velocities that may bias the fit when the uncertainty on the effective radius for a given galaxy is small.

Following \cite{Saglia2001} and \cite{Magoulas2012}, the scatter matrix $\boldsymbol{\Sigma}$ is decomposed into orthogonal unit eigenvectors $\boldsymbol{\hat{v}}_{1},\boldsymbol{\hat{v}}_{2},\boldsymbol{\hat{v}}_{3}$. These can be defined in terms of the FP parameters from the relationship in Eq~\ref{eq:FP_relation}. From these definitions, the Jacobian can then be used to write the terms in Eq.~\ref{eq:covariance} as functions of the FP parameters $a$ and $b$ and the intrinsic scatter along each orthogonal direction, $\sigma_{1}$, $\sigma_{2}$ and $\sigma_{3}$. The conversions between the two coordinate systems are given in Appendix~\ref{app:FPvectors}. In our fit we follow \cite{Saglia2001}, \cite{Colless2001} and \cite{Magoulas2012}, and assume \textit{a priori} that the longest axis of the 3D Gaussian lies exactly in the $r$-$i$ plane.

The best-fitting FP hence consists of a total of 8 free parameters: $a$, $b$, $\bar{r}$, $\bar{s}$, $\bar{\imath}$, $\sigma_{1}$, $\sigma_{2}$ and $\sigma_{3}$. We fit these by maximising the log of the likelihood function in Eq.~\ref{eq:like}. The maximisation is done using \textsc{scipy}'s implementation of the differential evolution optimisation algorithm \citep{Storn1997}, which provides robust fits in large, multidimensional parameter spaces without requiring gradients. The fitting algorithm can be made extremely fast (taking less than a minute to find the best fit on a single core) by utilising analytic computation wherever possible. This includes writing the determinant and inverse of the covariance matrix for each galaxy as a function of the matrix components rather than numerically inverting (trivial given this is only a $3\times3$ matrix), as this enables the calculation over all galaxies to be fully vectorised. The only remaining obstacle is the inclusion of selection effects, which as we will show in the next section can also be accounted for `analytically' and vectorised using elementary functions.

\subsection{Selection effects}

The SDSS PV sample has several selection effects that need to be accounted for both when fitting the FP and deriving log-distance ratios. These are:
\begin{itemize}\vspace*{-6pt}
    \item{lower and upper redshifts limit of $z_{\mathrm{min}}=0.0033$ and $z_{\mathrm{max}}=0.1$ respectively;}
    \item{a lower limit on velocity dispersions arising from the instrumental resolution of the SDSS spectrograph $s_{\mathrm{min}} = \mathrm{log}(70)$; and}
    \item{magnitudes limited to the range $10.0 \le m^{\mathrm{deV}}_{r} \le 17.0$.}
\end{itemize}

The likelihood function in Eq.~\ref{eq:like} can account for these selection functions via the normalisation $f_{n}$ and the $1/S_{n}$ weights. However, which of these are required and how they are computed depends on what exactly is being fit and so changes whether we are fitting the FP or using the fitted FP to extract peculiar velocities for each galaxy. In this section we will focus on the former; modelling of the selection function when fitting log-distance ratios will be tackled in Section~\ref{sec:logdist}.

\subsubsection{Magnitude and redshift limits}

Firstly, for a survey with both upper and lower redshift limits and a magnitude limit, like ours, we need to account for the fact that our observed sample is not a complete representation of the underlying FP from which the galaxies are drawn. At certain distances, galaxies will fall below the magnitude limit of the survey, which cuts a slice through the FP as a function of $r$ and $i$. To account for this, we upweight the galaxies that we \textit{have} observed using the $1/S_{n}$ factor. This is computed based on the fraction of the enclosed survey volume in which each galaxy with apparent magnitude $m^{\mathrm{deV}}_{r,n}$ could be observed,
\begin{equation}
    S_{n} =
    \begin{cases}
    1 & z_{\mathrm{lim}} \ge z_{\mathrm{max},n} \\
    \frac{d_{L}^{3}(z_{\mathrm{lim},n}) - d_{L}^{3}(z_{\mathrm{min}})}{d_{L}^{3}(z_{\mathrm{max}}) - d_{L}^{3}(z_{\mathrm{min}})} & z_{\mathrm{min}} < z_{\mathrm{lim},n} < z_{\mathrm{max}} \\
    0 & z_{\mathrm{lim},n} \le z_{\mathrm{min}}
    \end{cases}
\label{eq:Sn}
\end{equation}
where $d_{L}(z)$ is the luminosity distance to the redshift $z$ and $d_{L}(z_{\mathrm{lim},n})$ is the limiting distance for each galaxy. The latter is computed, given our $r$-band magnitude limit, using $d_{L}(z_{\mathrm{lim},n})=d_{L}(z_{\mathrm{group},n})\times10^{(17-m^{\mathrm{deV}}_{r,n})/5}$ where $z_{\mathrm{group},n}$ is the group-averaged redshift to galaxy $n$. The limiting redshift is then obtained from the limiting distance by inverting the redshift-distance relation. The form of the $S_{n}$ calculation is such that for galaxies bright enough that the limiting redshift is greater than the maximum redshift, we are complete and all possible galaxies at this magnitude have been included in our sample (modulo the $s_{\mathrm{min}}$ cut that will be discussed shortly); hence the weight is 1. For galaxies with limiting redshift below $z_{\mathrm{max}}$ we have only observed, on average, $S_{n}$ of these galaxies. We account for the impact of the missing galaxies on the FP fit by upweighting, taking the likelihood of the $n$th galaxy to the power $1/S_n$.

\subsubsection{Velocity dispersion limits}

As the $S_{n}$ weighting accounts for both the redshift and magnitude limits of our survey, only the velocity dispersion remains to be to dealt with. This is accounted for using the $f_{n}$ normalisation to rescale the PDF of each galaxy based on the volume of the FP parameter space that is not observed. Note that a similar approach could also have been utilised for the magnitude limits (and this is what is done when fitting the log-distance ratios). However, while this is the more principled approach (in a Bayesian sense), it is also far harder to implement than the $S_{n}$ weighting as it requires knowledge of the magnitudes \textit{and} log-distance ratios of galaxies that by definition have not been observed. This would require simulations for each set of FP parameters enclosed within a Bayesian hierarchical model. Such an approach is not necessary for modelling only the $s$ cut, as the impact of this cut on each galaxy does not depend on the galaxy's cosmological distance. It is also not required for fitting the log-distance ratio for the galaxies that \textit{have} successfully made it into our sample once the FP parameters are fixed.

Given the simplification that we only need to consider the limit imposed by $s_{\mathrm{min}}$, the computation of $f_{n}$ can be written as a 3D integral over the Gaussian in Eq.~\ref{eq:like}. Such an integral can be reduced to a single complementary error function as demonstrated in Appendix~\ref{sec:appfn2}, which depends primarily on the value of $s_{\mathrm{min}}$, and only weakly on the properties of each individual galaxy through the error matrix. This dependence is simple to understand; the portion of the FP missing from the observations is related to how far the velocity dispersion cut is from the mean of the sample given the scatter. A sample with mean far higher than the cut and small scatter will be effectively complete, whereas a sample with cut close to or higher than the mean will be heavily censored. For the SDSS PV sample, the velocity dispersion cut is considerably lower than the sample mean and the covariance is dominated by the intrinsic scatter in the FP, which is the same for all galaxies, rather than the measurement errors. Consequently, $f_{n}$ is similar, and close to unity, for most galaxies; $99\%$ of the SDSS PV sample have $f_{n} \ge 0.998$.

\subsection{Outlier Rejection}
\label{sec:FPoutliers}

Using the above methodology, we fit the FP iteratively in order to test the impact of outliers on the fitting procedure. At each iteration, we fit the FP, compute the $\chi^{2}$ difference between each data-point and the best-fitting model, then compute the $p$-value for each galaxy given the total log-likelihood of the fit and number of galaxies. Outliers are identified as galaxies with $p$ < $0.01$ and then excluded from the fit for the next iteration. The reduced $\chi^{2}$ is originally found to be $\sim$0.9, similar to that found in \cite{Magoulas2012}, which indicates that the best-fit intrinsic scatter is being overestimated to accommodate the outliers, but this quickly converges to $1.005$ over several iterations as outliers are removed. We find that the procedure converges after 5 iterations, with 503 outlier galaxies rejected from our sample (these are removed entirely from the PV catalogue), leaving us with a total of $34,059$.

\subsection{Results}
\label{sec:FPresults}

\begin{figure*}
\centering
\includegraphics[width=0.48\textwidth, trim=0pt 0pt 0pt 0pt, clip]{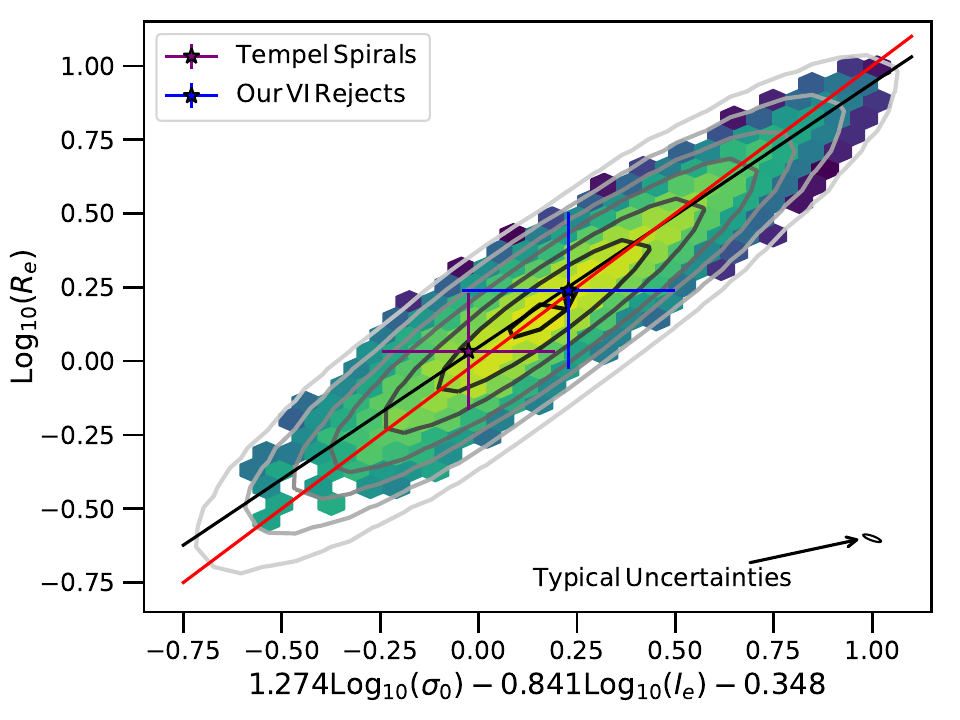}
\includegraphics[width=0.48\textwidth, trim=0pt 0pt 0pt 0pt, clip]{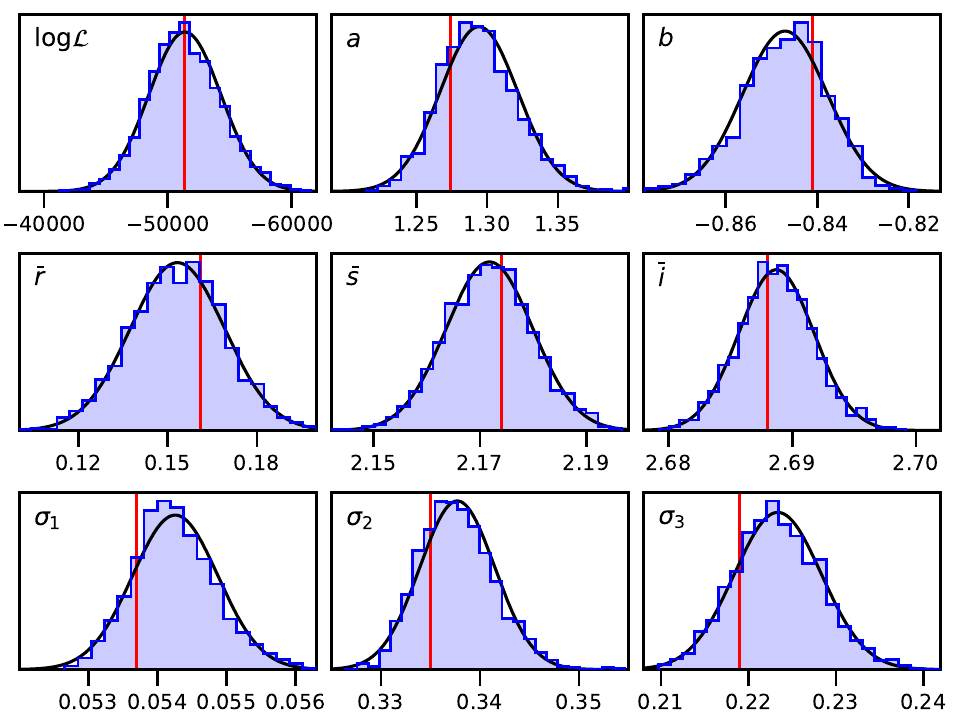}
  \caption{A comparison of the data and mock FP and best-fitting parameters. \textit{Left}: The measured effective radii against the predicted effective radii based on the best-fitting FP parameters for the SDSS data, weighted by the $1/S_{n}$ factor (Eq.~\ref{eq:Sn}). Coloured bins show the sum of the $1/S_{n}$ weighted data points, contours show the average distribution of the mocks. The impact of the $1/S_{n}$ weights is to up-weight galaxies with small effective radius that are generally faint enough they could have been missed at higher redshifts. The red line is the one-to-one line, which alongside the $\chi^{2}$ of our 3D Gaussian fit (102704 for 102169 degrees of freedom), demonstrates excellent agreement between the predicted and measured effective radii. The black line is the fit without including the $1/S_{n}$ weighting, which clearly demonstrates that the weighting is accounting for faint galaxies missing from our sample (which would have resided in the lower left corner of the FP). The blue and purple points show the mean and variance of the samples rejected by us, or because \protect\cite{Tempel2014} classify them as spirals, respectively. Finally, the small ellipse in the lower right shows the average (correlated) uncertainties for the data. \textit{Right:} The distribution of FP parameters measured from the mocks (blue histograms), overlaid with a Gaussian centred on the mean of mocks and with variance given as the uncertainty on the data in Table~\ref{tab:FP}. The top-left sub-panel is the log-likelihood for the best-fitting FP. The vertical red-lines are the best fit parameters for the data, which are consistent with the distribution of mock realisations.}
  \label{fig:SDSS_FP}
\end{figure*}

The final best-fit FP parameters for the SDSS data are given in Table~\ref{tab:FP}. In the same table we also provide FP fits for galaxies identified as spirals by \cite{Tempel2014} ($M == 1$) and/or rejected by us ($J == 0$). In each case we adopt our iterative outlier rejection method. These fits demonstrate that the rejected galaxies clearly populate distinct regions of the FP parameter space -- as identified in Section~\ref{sec:spirals}, the \cite{Tempel2014} spirals are typically more compact, being smaller with higher surface brightness, which is reflected in the drastic difference in $\bar{r}$ and $\bar{\imath}$ compared to our fiducial sample. The objects we reject are the opposite, being substantially larger in extent than our fiducial sample, and with smaller average surface brightness (although in such a way that they seem to shift along the plane in Fig.~\ref{fig:SDSS_FP}). Although comprising only $\sim$20$\%$ of the underlying sample, the rejected galaxies are enough to lead to a systematic shift in the FP fit when they are included compared to our fiducial case. This would lead to biased log-distance ratios for the rejects, \textit{and} for the clean early-types if they were left in.

\begin{table}
\centering
\caption{FP parameters for the SDSS PV sample and subsets. Columns $M==1$ and $J==0$ are subsets of spiral galaxies identified by \protect\cite{Tempel2014} and us, respectively.}
\begin{tabular}{lcccc} \hline
Parameter & Fiducial & $M == 1$ & $J == 0$ \\ \hline
$N_{\mathrm{gal}}$ & 34,059 & 5,033 & 2,931 \vspace{1pt} \\
$a$ & $1.274 \pm 0.027$ & 1.153 & 1.296  \\
$b$ & $-0.841 \pm 0.009$ & $-$0.738 & $-$0.821 \\
$\bar{r}$ & $0.161 \pm 0.016$ & 0.027 & 0.250  \\
$\bar{s}$ & $2.174 \pm 0.008$ & 2.154 & 2.198 \\
$\bar{i}$& $2.688 \pm 0.003$ & 2.893 & 2.638  \\
$\sigma_{1}$ & $0.0537 \pm 0.0006$ & 0.0472 & 0.0517 \\
$\sigma_{2}$ & $0.335 \pm 0.004$ & 0.315 & 0.367 \\
$\sigma_{3}$ & $0.219 \pm 0.005$ & 0.191 & 0.179
\vspace{1pt} \\ \hline
\end{tabular}
\label{tab:FP}
\end{table}
 
We also apply the same fitting procedure to all of our mock galaxy catalogues. This allows us to test how well our simulations match the distribution of the data, and also estimate uncertainties on our fiducial sample using the standard deviations between mock realisations. The distribution of mock FP parameters in shown in Fig~\ref{fig:SDSS_FP}. We see that the mocks match the fit from the data extremely well.

Overall, the key assumption in the above fitting methodology is that the data is well described by a 3D Gaussian. Although we do not plot them here, looking at the individual distributions of $r$ and $i$ provided with the publicly released data file, one can see that this is clearly true. There is some skewness in $s$, with a slightly elongated tail of low velocity dispersion galaxies. Nonetheless, the small number of outliers removed during our fitting procedure, chi-squared per degree of freedom very close to one and log-likelihood that is well-reproduced by mocks that \textit{are}, by construction, 3D Gaussian distributed, demonstrates that this skewness is not unduly affecting our fits and the model works well for the SDSS PV data. We do caution however that this may not be the case for future, larger datasets.

\section{Fitting the log-distance ratios}
\label{sec:logdist}

Given the relationship between log-distance ratio and the difference in effective sizes described in Section~\ref{sec:FPtoPV}, one can then use the probability distribution of the FP given in Eq.~\ref{eq:like} to fit the log-distance ratio of each galaxy by using the modified set of variables $\boldsymbol{x}_{n} = \{r_{n}-\eta_{n}$, $s_{n}$, $i_{n}\}$ and fixing $\boldsymbol{\bar{x}}$ and $\boldsymbol{\mathsf{C}}_{n}$ based on the best-fit FP parameters and observational uncertainties for each galaxy.

\begin{figure}
\centering
\includegraphics[width=0.48\textwidth, trim=0pt 0pt 0pt 0pt, clip]{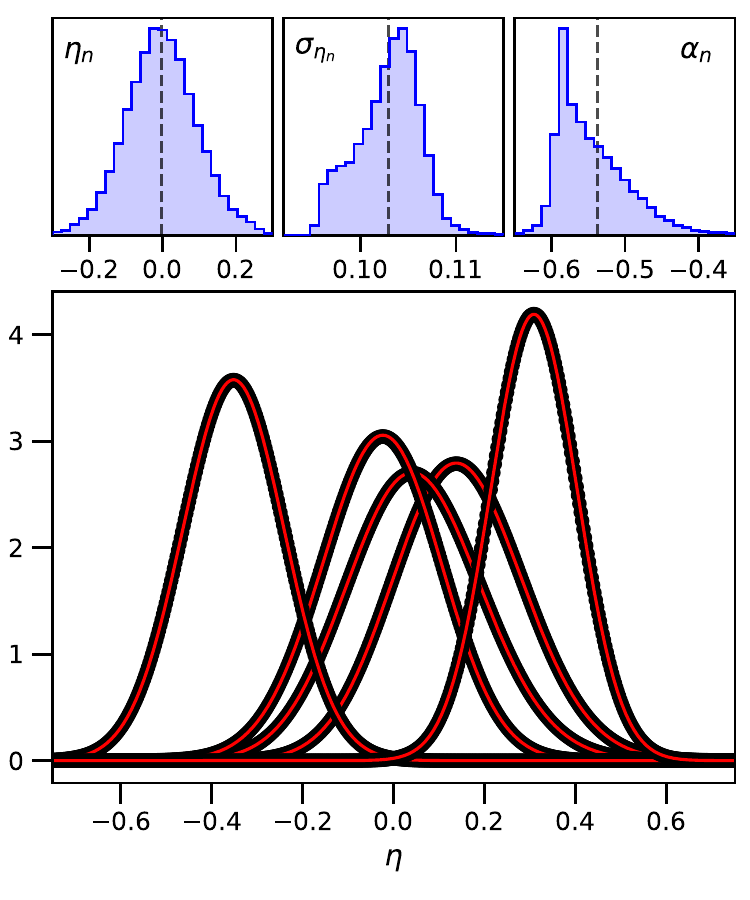}
  \caption{\textit{Top}: The distribution of means, standard deviations and skewness parameters for the SDSS PV data. \textit{Bottom:} Example normalised probability distributions for the log-distance ratio of galaxies in the SDSS PV sample. We purposely plot the galaxies with the largest/smallest mean and skewness, and the largest uncertainty. In all cases, the skew-normal provides an excellent fit (red line), and the distributions are close to Gaussian. Note that these distributions include the $f_{n}$ correction described in Section~\ref{sec:fn} and are plotted with $2.5\times$ fewer points than is actually used in our fitting. }
  \label{fig:logdists}
\end{figure}

To `fit' the log-distance ratio, we generate 1001 uniformly distributed candidate values for the log-distance ratio of each galaxy in the range [$-$1.5,1.5] and compute the log-likelihood for each. We then combine this with a flat prior on the log-distance ratio and normalise to obtain a finely tabulated posterior PDF for each galaxy, $P(\eta_{n} | r_{n}$, $s_{n}$, $i_{n}$, $\boldsymbol{\bar{x}}$, $\boldsymbol{\mathsf{C}}_{n})$. Summary statistics are then produced by assuming the posterior PDF of each galaxy can be represented by a skew-normal distribution \citep{OHagan1976,Azzalini1985} with location ($\xi_{n}$), scale ($\omega_{n}$) and shape ($\alpha_{n}$) parameters
\begin{multline}
P(\eta_{n} | r_{n}, s_{n}, i_{n}, \boldsymbol{\bar{x}}, \boldsymbol{\mathsf{C}}_{n}) = \frac{1}{\sqrt{2\pi}\omega_{n}}\exp\biggl[-\frac{(\eta_{n}-\xi_{n})^{2}}{2\omega^{2}_{n}}\biggl] \\
\biggl(1+\mathrm{erf}\biggl[\alpha_{n}\frac{\eta_{n}-\xi_{n}}{\sqrt{2}\omega_{n}}\biggl]\biggl) ~,
\label{eq:skewnormal}
\end{multline}
where $\mathrm{erf}(z)$ is the error function. We find that this distribution provides an excellent representation of the SDSS galaxy posteriors, and captures the small skew in the log-distance ratio PDFs that arises from the $f_{n}$ correction for the selection function described below. Example PDFs and corresponding skew-normal distributions are shown in Figure~\ref{fig:logdists}, where we purposely plot the objects with the largest/smallest skewness and mean, and the largest uncertainty. A similar distribution was used in \cite{Springob2014}. The location, scale, and shape parameters for each galaxy can be estimated from the mean ($\langle \eta_{n} \rangle$), standard deviation ($\sigma_{\eta_{n}}$) and skewness ($\gamma_{n}$) of the tabulated posteriors using the following relations,
\begin{align}
\xi_{n}    &= \langle \eta_n \rangle - \omega_{n}\delta_{n}\sqrt{\frac{2}{\pi}} ~, \quad \omega_{n} = \sigma_{\eta_n}\sqrt{\frac{\pi}{\pi-2\delta_n^{2}}} ~, \nonumber \\
\alpha_{n} &= \frac{\delta_n}{\sqrt{1 -\delta^{2}_{n}}} ~, \quad |\delta_{n}| = \sqrt{ \frac{\pi|\gamma_{n}|^{2/3}}{2|\gamma_{n}|^{2/3} + (\sqrt{2}(4-\pi))^{2/3}}} ~.
\end{align}
When $\alpha_{n}$\,=\,0, Eq.~\ref{eq:skewnormal} reduces to a normal distribution with the mean and standard deviation used above. Though the $\alpha_{n}$ values shown in Figure~\ref{fig:logdists} are consistently non-zero, the skewness of the SDSS galaxies is small enough that it will likely have a negligible effect on subsequent analyses of our catalogue, but we recommend confirming this where possible.

When evaluating the log-likelihood as a function of $\eta$ there are two terms that complicate matters: a multiplication by $1/S_{n}$ and the addition of $-\ln(f_{n})$. Unlike the case when fitting the FP, we can now ignore the $1/S_{n}$ weight because it is fixed for a given galaxy; however we cannot ignore the $f_{n}$ term. 

\subsection{\texorpdfstring{$\mathbf{f_{n}}$}{} correction}
\label{sec:fn}

$f_{n}$ describes the normalisation of each galaxy's PDF and mathematically encapsulates the `selection bias' in our sample. Hence it must be accurately computed for each galaxy and each possible distance to that galaxy. The effect of including $f_{n}$ is to up-weight larger distances for each galaxy to account for (1)~the sampling of the FP being less complete at large distances where fainter galaxies drop below our magnitude limit and (2)~observed galaxies being more likely at larger distances than smaller distances because there is simply more volume at larger distances.

Mathematically, $f_{n}$ is the integral of the Gaussian likelihood, but with limits imposed by our selection function. These are a lower limit for velocity dispersions, $s>\log(70)$, which doesn't depend on distance and so actually normalises out of our fitting anyway, and a limited range in magnitude, $10.0 \le m^{\mathrm{deV}}_{r} \le 17.0$. The latter cuts through the FP in a way that can be written in terms of the minimum and maximum effective sizes given a value for the surface brightness. This can be seen by rearranging the relationship between apparent magnitude, effective radius and surface brightness in Eqs.~\ref{eq:FPr} and ~\ref{eq:FPi}, and substituting in the magnitude limits. Hence,
\begin{multline}
    f_{n} = \frac{1}{(2\pi)^{3/2}|\boldsymbol{\mathsf{C}}_{n}|^{1/2}}\int_{-\infty}^{\infty}\int_{r_{\mathrm{min}}-i/2}^{r_{\mathrm{max}}-i/2}\int_{s_{\mathrm{min}}}^{\infty}ds dr di \\ \exp\biggl\{-\frac{1}{2}(\boldsymbol{x_{n}}-\boldsymbol{\bar{x}})\boldsymbol{\mathsf{C}}_{n}^{-1}(\boldsymbol{x_{n}}-\boldsymbol{\bar{x}})^{T}\biggl\}
\end{multline} 
where
\begin{align}
    r_{\mathrm{min}} = & (10 + M^{r}_{\odot} + 5\log(1+z_{\mathrm{helio}}) - 0.85z_{\mathrm{group}} \notag \\ 
    & - 2.5\log(2\pi) + k_{r} + A_{r} + 5\log(d(\bar{z})) - 17)/5 ~, \\
    r_{\mathrm{max}} = & (10 + M^{r}_{\odot} + 5\log(1+z_{\mathrm{helio}}) - 0.85z_{\mathrm{group}} \notag \\ 
    & - 2.5\log(2\pi) + k_{r} + A_{r} + 5\log(d(\bar{z})) - 10)/5 ~,
\end{align}
and both the cosmological redshift $\bar{z}$ and the comoving distance to that redshift, $d(\bar{z})$, are computed based on each galaxy's observed redshift and the candidate log-distance ratios. Hence there are actually 1001 values of $f_{n}$ for each galaxy, varying as the proposed distance to the galaxy increases. The individuality of the $f_{n}$ for each galaxy in the SDSS sample is also apparent in the fact that the $k$-correction, extinction and covariance matrix that enter into the above equations vary from galaxy-to-galaxy.

The large number of galaxies in an FP sample means that evaluating the above integral becomes computationally demanding if one were to use numerical integration. To circumvent this, \cite{Springob2014} used simulations before and after the selection functions were applied as a form of Monte Carlo integration and assumed the same correction for each galaxy. Besides this assumption, the disadvantage of this method is that the $f_{n}$ calculation remains quite inaccurate at large distances, where, by construction, the number of mock galaxies with which we can compute $f_{n}$ quickly goes to zero, even when large samples are generated. In Appendix~\ref{sec:appfn1}, we show that, although complex, the above integral can actually be reduced to a sum of elementary functions that are fast to evaluate using standard computational libraries (such as \textsc{scipy} in Python) and give an exact solution. As a result, we are able to compute $f_{n}$ for every galaxy in the SDSS sample at each of their 1001 candidate distances in less than a minute. 

\begin{figure}
\centering
\includegraphics[width=0.48\textwidth, trim=0pt 0pt 0pt 0pt, clip]{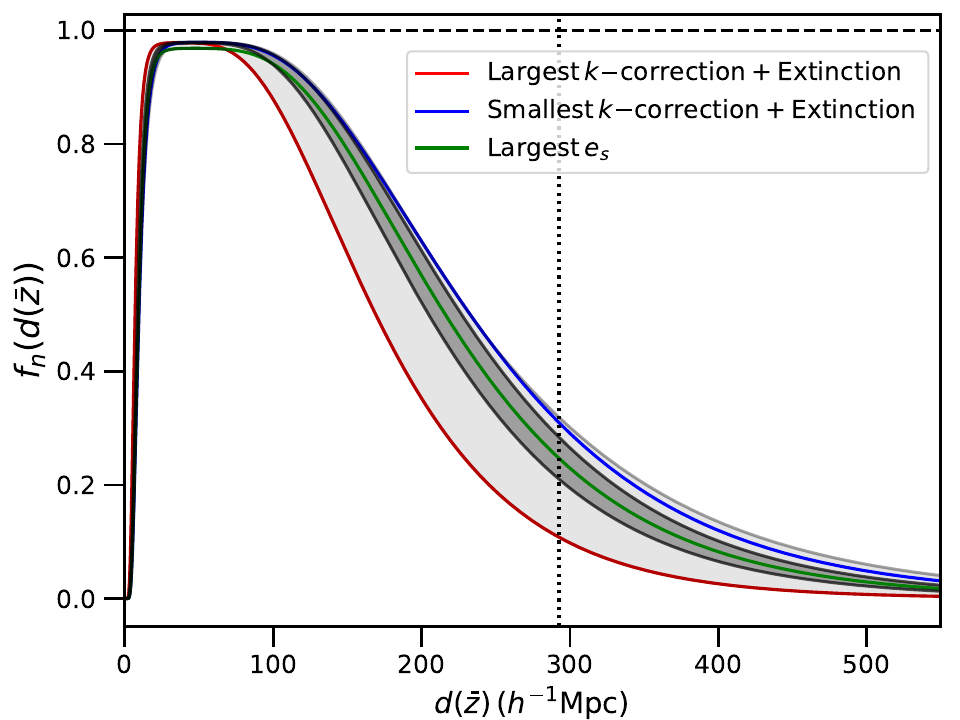}
  \caption{The $f_{n}$ normalisation for each galaxy in the SDSS sample as a function of proposed distance. The vertical dotted line shows the comoving distance to the maximum redshift ($z=0.1$) of our sample assuming our fiducial cosmology. The light grey area shows the region in which \textit{all} galaxies reside, while the dark grey region contains $95\%$ of the galaxies. In addition, curves for three individual galaxies are shown to highlight the typical shape of the curve and also the sources of variation in the relationship between galaxies. Galaxies with smaller $k$-corrections and extinctions have $f_{n}$ relations typically shifted to greater distances compared to those with larger values; if they both had the same effective size and surface brightness, the one with the smaller sum of $k$-correction and extinction would be observable to a larger distance. In addition, galaxies with larger velocity dispersion uncertainties ($e_s$) are more likely to have scattered into the sample from below the velocity dispersion cut.}
  \label{fig:fn}
\end{figure}

The resulting function is shown in Figure~\ref{fig:fn}. The characteristic shape of $f_{n}$ as a function of distance is a peak lying in the range 10--100\,$h^{-1}$\,Mpc with a value close to unity (but not exactly, due to the $s_{\mathrm{min}}$ cut) that drops towards zero at very small distances and at larger distances due to the bright and faint magnitude limits respectively. For about 95\% of galaxies the curves are very similar, which validates the assumption made by \cite{Springob2014}; however there are some outlying galaxies with more extreme distributions as a function of distance. Broadly speaking, the combination of $k$-correction and extinction that enters into the conversion from apparent magnitude to effective radius and surface brightness varies the distance scale of the $f_{n}$ relationship; two galaxies with the same effective radius and surface brightness but different colours or at different proximity to the Galactic plane may not be observable to the same distance. The other trend seen in Figure~\ref{fig:fn} is related to the galaxy's velocity dispersion uncertainty. Although all galaxies are assumed to be drawn from the same best-fitting FP and are subject to the same $s_{\mathrm{min}}$ cut, galaxies with larger velocity dispersion uncertainties were more likely to have scattered into the sample erroneously, and so, at its peak, the $f_{n}$ value is slightly further from unity.

\subsection{Tests on mock catalogues and residual bias corrections}

Given the method for estimating the log-distance ratio in the presence of selection effects described above, we turn to validating how well we recover the true log-distance ratios in our mocks. We test the extent of residual biases needing to be corrected in the data. This also allows us to assign realistic measured values and errors to the mock data, which will be useful for computing (for instance) the uncertainty on cosmological parameters measured from the SDSS PV sample.

\subsubsection{Trends as a function of redshift}
\label{sec:red_bias}

\begin{figure}
\centering
\includegraphics[width=0.48\textwidth, trim=0pt 0pt 0pt 0pt, clip]{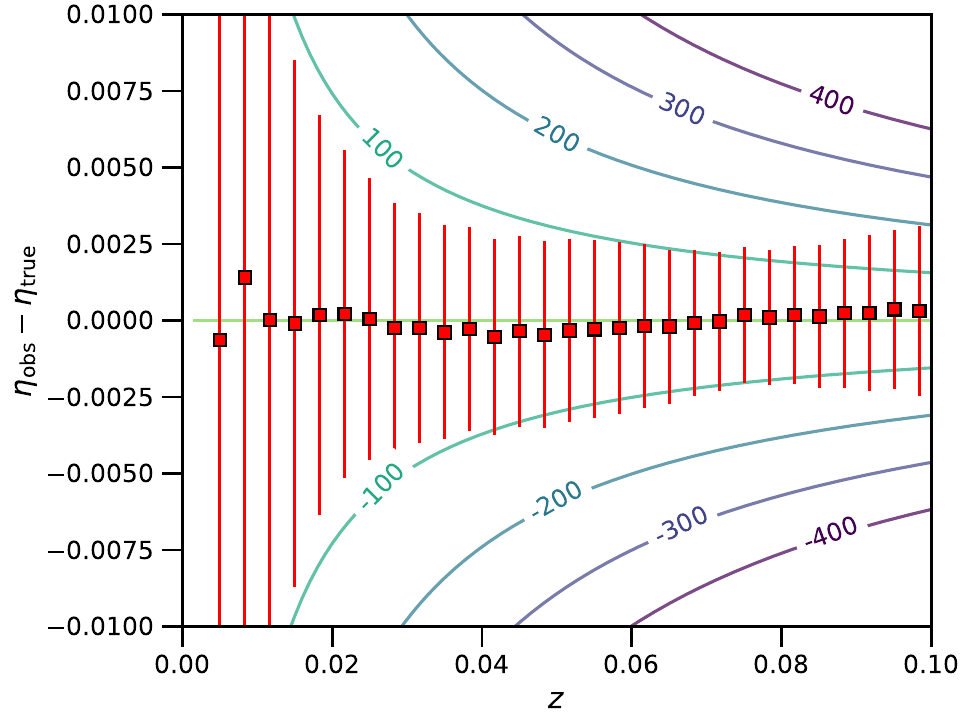}
  \caption{Bias in the measured log-distance ratios from SDSS PV mocks, binned as a function of redshift. Each data point corresponds to the error-weighted mean, while the error bar represents the standard deviation of the mocks in each bin (\textit{not} the standard error of the mean, which would be $\sqrt{2048}$ times smaller). Lines of constant peculiar velocity are also shown. Our pipeline produces measured log-distance ratios that are unbiased across the entire redshift range of our sample, and in excellent agreement with the truth values from the simulations.}
  \label{fig:red_bias}
\end{figure}

We first examine the difference between the true and measured log-distance ratios ($\eta_{\mathrm{true}}$ and $\eta$ respectively) in the mocks as a function of redshift. A bias as a function of redshift translates into a spurious inflow/outflow, and even a small offset in log-distance ratio can lead to a large bias in peculiar velocity at the high redshift end of our data. Figure~\ref{fig:red_bias} shows that our methodology for first fitting the FP, then extracting log-distance ratios, when applied to all our mocks, produces results that are unbiased. There is excellent agreement between the measured and true log-distance ratios averaged over the 2048 simulations, and no evidence that our pipeline is introducing spurious flows.

\subsubsection{Trends as a function of magnitude}
\label{sec:magbias}

\begin{figure}
\centering
\includegraphics[width=0.48\textwidth, trim=10pt 0pt 0pt 0pt, clip]{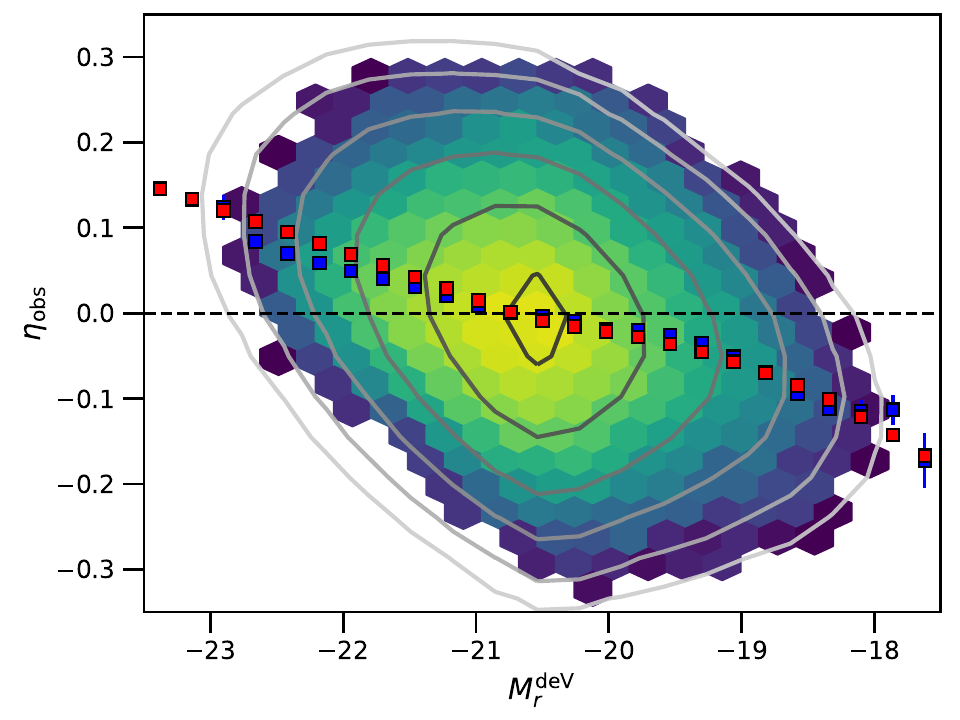}
  \caption{A plot of the trend between absolute $r$-band magnitude and log-distance ratio seen in our mocks and data. Hexbins show real galaxies, with brighter colours indicating a higher density; contours show mock galaxies, with darker contours indicating a higher density. Red (blue) points are the average log-distance ratio of the mocks (data) in bins of absolute magnitude. The log-distance ratios clearly show a trend with absolute magnitude, but this does not lead to any biases.}
  \label{fig:magbias}
\end{figure}

Further investigation into the mock catalogues and data reveals a trend between the absolute magnitude and the recovered log-distance ratios. In both the mock catalogues and the data, intrinsically bright (faint) galaxies have log-distance ratios that are systematically higher (lower) than the mean. This translates into a similar trend as function of apparent magnitude, and  is shown in Figure~\ref{fig:magbias}. 

Although at first glance this is a concern, we have verified that this trend is an expected result of identifying the best-fit log-distance ratio from the offset between $r$ and the 3D Gaussian FP. This can be understood by considering that, from Eqs.~\ref{eq:FPr} and~\ref{eq:FPi}, the absolute magnitude of a galaxy $M^{\mathrm{deV}}_{r} \propto r + 0.5i$.  However, the maximum likelihood log-distance ratio is given by the offset from the FP in the $r$-direction, which from Eq.~\ref{eq:FP_relation} (ignoring the $f_{n}$ term and other complexities) means $\eta \propto r + 0.841i$ for the SDSS PV sample (which has $b=-0.841$; see Table~\ref{tab:FP}). Putting these together clearly shows that we expect $\eta \propto M^{\mathrm{deV}}_{r}$, albeit with some residual dependence on the surface brightness and velocity dispersion. 

\begin{figure}
\centering
\includegraphics[width=0.48\textwidth, trim=0pt 0pt 0pt 0pt, clip]{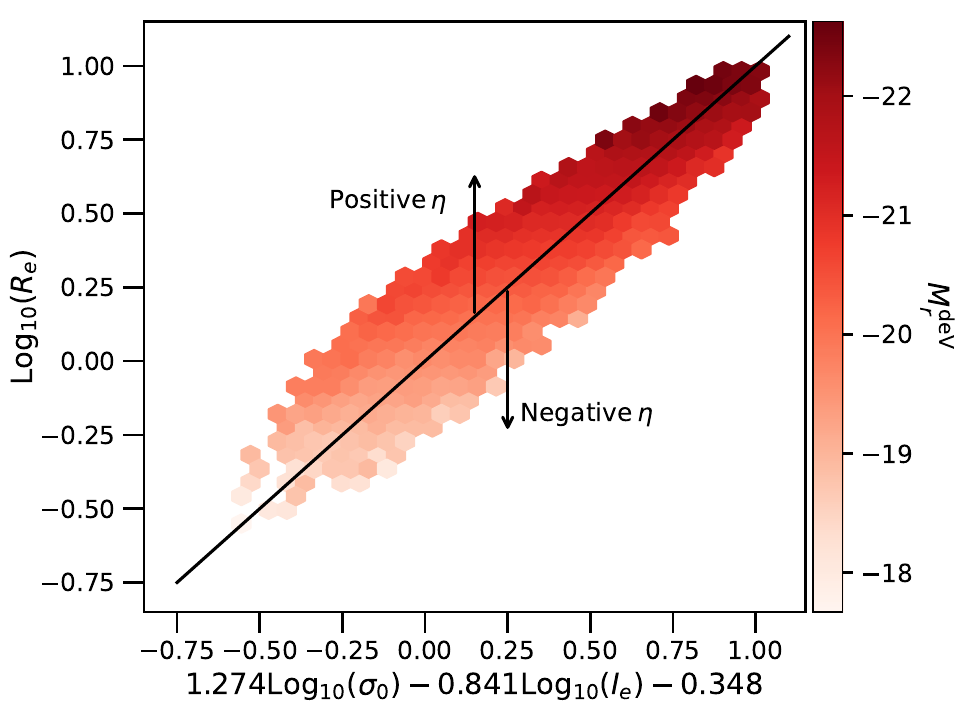}
  \caption{The measured effective radii against the predicted effective radii based on the best-fitting FP parameters for the SDSS data, as per Fig.~\ref{fig:SDSS_FP}, but with bins coloured according to the mean absolute magnitude. The black line shows the one-to-one line, and the vertical arrows show how the recovered log-distance ratio is related to this projection of the FP. One can see that there is a correlation between absolute magnitude and position above/below the FP (rather than just along it) that inevitably creates the observed trend between recovered log-distance ratio and absolute magnitude.}
\label{fig:FP_Mag}
\end{figure}

This can be seen graphically in Figure~\ref{fig:FP_Mag}, where we plot the SDSS FP data as in Figure~\ref{fig:SDSS_FP}, but with each bin colour-coded by the average absolute magnitude. Although subtle, there is a preference for intrinsically brighter/fainter galaxies to be situated above/below the plane, which then results in the brighter/fainter galaxies having positive/negative log-distance ratios, exactly as discussed mathematically above, and seen in Figure~\ref{fig:magbias}. This trend might be diminished if one were to use an alternate method of fitting the FP and extracting log-distance ratios --- however, \cite{Saglia2001} explored a number of these alternatives and found that the Maximum likelihood 3D Gaussian we use here gives the most unbiased peculiar velocities, despite the apparent trend with absolute magnitude, as it more accurately accounts for the range of sizes, velocity dispersions and surface brightnesses seen in a typical FP sample as well as simultaneous (and potentially correlated) errors in all three parameters.

We did investigate whether this trend could lead to biases in subsequent uses of the SDSS PV catalogue and should be corrected. We concluded that this trend does not result in a bias because: (i)~we have already demonstrated that the mocks are unbiased as a function of redshift (Figure~\ref{fig:red_bias}), even though there is a trend with absolute magnitude; (ii)~our sample is almost volume limited up to even large distances --- at $z=0.05$, corresponding to a comoving distance of $150h^{-1}\,\mathrm{Mpc}$, the limiting absolute magnitude is only $M_{r}^{\mathrm{deV}}\le -18.9$; (iii)~our tests of the bulk flow and growth rate measurements obtained from the mocks (presented here in Section~\ref{sec:cosmo} and in Lai et~al., in prep., respectively) show no significant biases in the recovered measurements; and (iv)~as discussed in the next section, we do find a bias associated with group richness, that could conceivably have been due to groups containing galaxies that do not span the full range of absolute magnitudes. However, we implemented a correction forcing the log-distance ratio in the data and mocks to be flat as a function of absolute magnitude, and found it had \textit{no} effect on the bias with group richness (which is therefore corrected differently). Hence, we do not `correct' for the trend between log-distance ratio and absolute magnitude seen here. 

It is important to note that this could cause bias in subsequent analyses if the data were later cut to a brighter magnitude limit and averaged over because the FP and $f_{n}$ correction for all galaxies in the SDSS sample have been fit/computed assuming a magnitude limit of $17.0$. If the data is cut to a brighter limit, then the correct procedure would be to refit the FP and log-distance ratios using the updated magnitude limit.

\subsection{Trends as a function of group richness}
\label{sec:groupcorr}

In the previous sections, we have demonstrated that we are able to recover log-distance ratios that are unbiased as a function of redshift, and the observed trend with absolute magnitude is as expected and does not need correction. However, we do observe one source of systematic bias in the SDSS sample that does require correction: a correlation between the recovered log-distance ratio and the number of galaxies in the same group. We verified this is uncorrelated with the trend with absolute magnitude (i.e., due to the fact that groups with more members may contain different distributions of bright or faint galaxies), since forcing the average observed log-distance ratio to be constant as a function of absolute magnitude did not remove the bias with group richness.

\begin{figure}
\centering
\includegraphics[width=0.48\textwidth, trim=0pt 0pt 0pt 0pt, clip]{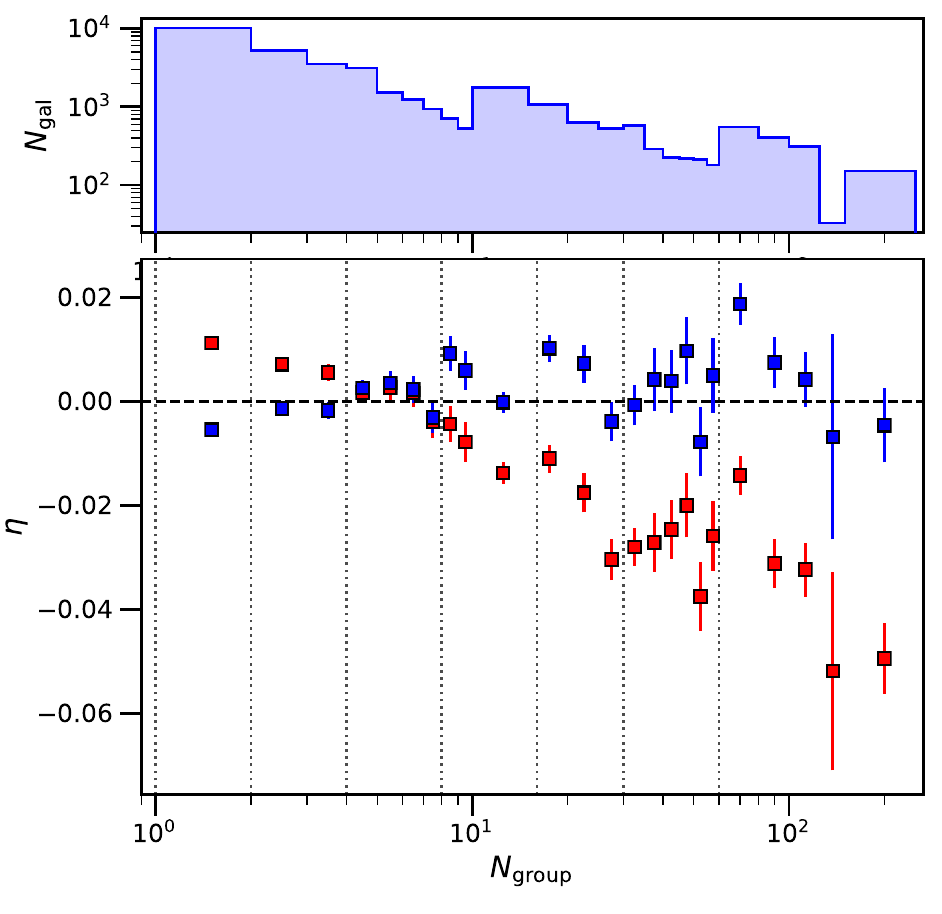}
  \caption{\textit{Top}: Number of SDSS PV galaxies as a function of group richness. \textit{Bottom}: Average log-distance ratio in the SDSS PV catalogue as a function of \protect\cite{Tempel2017} group richness. Red points show the log-distance ratios recovered using our default pipeline where all data is fit using a single FP, and that single set of FP parameters is used to recover log-distance ratios for the full sample. Blue points show the log-distance ratios when separate FPs are fit and used for different subsamples based on group richness as discussed in Section~\ref{sec:groupcorr} and denoted by the vertical dotted lines.} 
  \label{fig:logdist_ngroup}
\end{figure}

The bias is demonstrated in Fig.~\ref{fig:logdist_ngroup}, where we plot the average log-distance ratio over groups of different sizes, where for size we use the number of galaxies (not all of which are in the SDSS PV sample) belonging to the same \cite{Tempel2017} group. We see a clear trend of decreasing log-distance ratio with increasing group richness. This is a problem because larger groups will have more measured peculiar velocities, and it is common to average over the peculiar velocities within each group, exacerbating this bias.

The existence and origin of environmental dependencies in the FP, either in terms of local properties such as the distance from the galaxy to the cluster centre, or global properties such as cluster richness or radius, is a long-standing question. Previous studies using large samples of galaxies such as those of \cite{Bernardi2003c, Donofrio2008, LaBarbera2010, Magoulas2012} and \cite{Hou2015} find correlations between the FP offset/residuals with local surface density, but less evidence of correlations with group richness---although large differences are seen between galaxies in groups (of any size) and in the field. One possibility is that the observed change in the FP with cluster richness is the result of a more elementary correlation between the FP and the stellar age of a galaxy, with richer clusters containing more evolved stellar populations (d'Eugenio et. al., in prep). 

Another likely possibility is that the trend arises due to data systematics. There are a number of other trends in the data correlated with group richness; including redshift, apparent magnitude and angular size. However, a substantial fraction of this is to be expected --- for a magnitude limited sample, richer groups contain, on average, fainter galaxies which are found at lower redshifts with larger angular sizes on the sky. This effect should be partially compensated for by our accounting of the selection function when fitting the FP and log-distance ratios, and we see no bias in our log-distance ratios with redshift (which would clearly translate into a bias on group richness if this also varies with redshift). The bias remains if we remove all use of group redshifts in our FP data and fitting, which indicates it is not due to misidentification of, or misassociation with, groups.

It is hence difficult to disentangle what proportion of the observed trends are due to our sample being magnitude limited, data systematics or environmental effects, and we do not address this further in the current work. Nonetheless, we seek a way to ensure this does not lead to biased peculiar velocities, which is most easily achieved by fitting the FP to different subsamples of our full dataset based on the group richness.

We do this by splitting our sample into roughly logarithmic bins in group richness, where each subsample contains at least 1,500 galaxies and 70 distinct groups. We then run the separate subsamples through our default fitting methodology, recovering both the best-fit FP parameters and new log-distance ratios for each galaxy.

\begin{figure}
\centering
\includegraphics[width=0.48\textwidth, trim=0pt 0pt 0pt 0pt, clip]{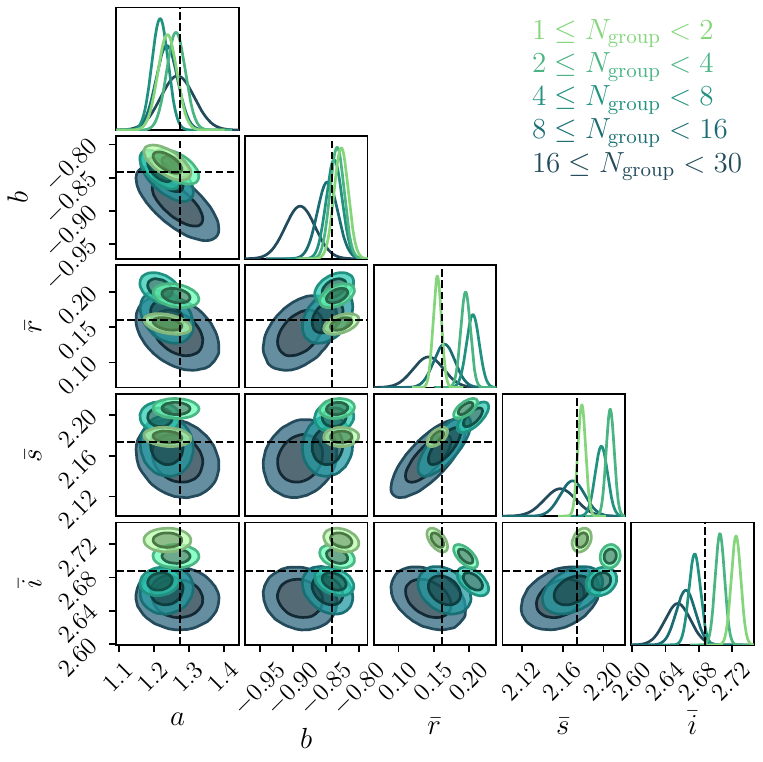}
  \caption{Distributions of FP parameters for subsamples of the SDSS PV catalogue split as a function of group richness. Contours and histograms are derived from 200 simulations centred on the best-fit parameters from the data, each with the same number of simulated data points as the real data subsamples. These are used to demonstrate the expected spread in FP parameters for each subsample. The black-dashed lines show the FP parameters fit to the full SDSS PV sample. There is considerable scatter in the best-fit parameters with group richness, but the slope $b$ and mean surface brightness $\bar{\imath}$ exhibit clear trends.} 
  \label{fig:groupfits}
\end{figure}

The FP parameters for the most constraining sub-samples along with errors derived from 200 simulations centred on the fits from the data are shown in Fig.~\ref{fig:groupfits}. The simulations are generated as for the single FP sample in Section~\ref{sec:mocks}, but without populating an underlying N-body simulation (and so do not have any large-scale structure or clustering). We also do not show the constraints for group sizes larger than 30 as the parameter constraints are too weak to deduce anything meaningful. Nonetheless, there are clear systematic variations of the slope $b$ and mean surface brightness $\bar{\imath}$, both of which decrease with increasing group richness. Correlations with other parameters are less clear. The trends with $b$ and $\bar{\imath}$ seem to be detected at high significance, but for other parameters the field population does not seem statistically different from the different size groups.

The origin of this remains unclear---the trend identified by \cite{LaBarbera2010} is with the parameter combination $c=\bar{r}-a\bar{s}-b\bar{\imath}$ and not reproduced here, while \cite{Magoulas2012} did not see a strong correlation between $b$ and group richness, but do with local galaxy surface density. However, we find that performing separate fits by group richness and then combining the resulting subsamples leads to log-distance ratios that exhibit far less bias with group richness. This is shown in Figure~\ref{fig:logdist_ngroup} as the blue points. 

As we are able to remove the bias with group richness by simply fitting separate FPs to the subsamples of the data, we use this as our empirical correction going forward. The `corrected' log-distance ratios provided with the SDSS PV catalogue are those obtained using these multiple FP fits rather than a single fit to the entire sample.

\subsection{Zero-point}
\label{sec:zp}

A key assumption when fitting the FP is that the net velocity of the sample is zero. This is unlikely to be true in reality. To correct for this assumption, we need to make one final `zero-point' correction to our sample. In \cite{Springob2014}, this was done by assuming that a sample drawn from an approximate great circle (in this case close to the celestial equator) truly does have net velocity equal to zero. However, this makes use of the hemispherical sky-coverage afforded by 6dFGSv, which is not available with the smaller footprint of the SDSS PV sample. Instead we calculate the zero-point of the SDSS PV sample by cross-matching to overlapping galaxies that also contain distance measurements in the Cosmicflows-III catalogue (CF3; \citealt{Tully2016}), using the individual redshifts to convert from distance moduli presented in CF3 to log-distance ratio. CF3 itself is calibrated using a distance ladder containing first galaxies hosting Cepheid variable stars \citep{Freedman2012}, Tip of the Red Giant Branch stars \citep{Rizzi2007} and maser emission \citep{Humphreys2013}; then Type Ia Supernovae \citep{Rest2014}. By linking the SDSS PV sample to CF3 we are hence extending this distance ladder and relying on the calibration of the intermediate rungs to set our zero-point. This also means that our result for the bulk flow in Section~\ref{sec:cosmo} will be strongly correlated with the same measurement from CF3. However, we do validate our zero-point using independent data from the 2M++ reconstruction of the local velocity field.

We perform the calibration in two ways, careful to fairly compare log-distance ratios before and after the correction for group richness in Section~\ref{sec:groupcorr}. We first look at the 296 individual galaxies in common between CF3 and the SDSS PV sample, of which 8, 6, 2 and 285 have previous distances from Type Ia supernovae, surface brightness fluctuations, the Tully-Fisher relation and the FP respectively.\footnote{This adds up to more than 296 galaxies as some galaxies have measurements using multiple tracers. In this case the individual CF3 measurements are averaged into a single value for the galaxy.} 
The CF3 FP-based distances were derived from the EFAR, SMAC, ENEAR and 6dFGSv surveys.
The presence of SDSS PV measurements that also have TF distances is worrying, but both of the above cases are lenticular galaxies where it is unclear from the visual inspection whether the FP or Tully-Fisher relation (or indeed both!) is more appropriate. To be conservative, we remove these from the zero-point calibration, but find that doing so changes the zero-point by only $0.1\sigma$. The remain 294 overlapping galaxies are distributed across the full redshift range of the SDSS PV sample, however the calibration is dominated by low redshift objects---$90\%$ have $z<0.05$.

From the 294 galaxies we compute the weighted mean difference between CF3 and log-distance ratios \textit{from our single FP fit to the full sample} (i.e., with no correction for group richness) as
\begin{equation}
    \langle \eta_{\mathrm{CF3}} - \eta_{\mathrm{SDSS}} \rangle = -0.0028\pm0.0080,
\end{equation}

The second way we compute the zero-point is using groups that share both a CF3 and SDSS PV measurement. Using the \citep{Tempel2017} group catalogue, we identify 292 groups that contain at least one CF3 and one SDSS PV measurement. We then first average the measurements within the two catalogues (i.e., if a group contains two CF3 and four SDSS PV measurements, we average the two to obtain a single CF3 consensus value, and the four to obtain a single SDSS PV consensus value). We then compare the difference at the group-averaged level between CF3 and our log-distance ratios obtained \textit{from multiple FPs fit to the full sample as a function of group richness} (i.e., correcting for the bias in Section~\ref{sec:groupcorr}). We find
\begin{equation}
    \langle \eta_{\mathrm{CF3}} - \eta^{\mathrm{corr}}_{\mathrm{SDSS}} \rangle = -0.0037\pm0.0040.
\label{eq:zeropoint}
\end{equation}
This is fully consistent with the zero-point from individual objects. However, the use of group-averages provides a smaller uncertainty, and so we adopt this as the official zero-point for the SDSS PV sample. As a final check, we predict the velocities for each of the SDSS PV galaxies using the SDSS PV redshifts and reconstructed velocity field of 2M++ (\citealt{Carrick2015}; processed as in \citealt{Carr2021}). We then convert these to log-distance ratios and evaluate the zero-point. For the single FP fit and multiple FP fit log-distance ratios we find
\begin{align}
    \langle \eta_{\mathrm{2M++}} - \eta_{\mathrm{SDSS}} \rangle &= -0.0019\pm0.0006, \\
    \langle \eta_{\mathrm{2M++}} - \eta^{\mathrm{corr}}_{\mathrm{SDSS}} \rangle &= -0.0013\pm0.0005,
\end{align}
respectively, which are both also consistent with our other methods of determining the zero-point. We do caution that this last method is not fully model independent - the predicted velocities of \cite{Carrick2015} depend on cosmological parameters - and so this is used only as a cross-check of the empirical zero-point (Eq.~\ref{eq:zeropoint}) we actually adopt.

\begin{figure}
\centering
\includegraphics[width=0.48\textwidth, trim=0pt 0pt 0pt 0pt, clip]{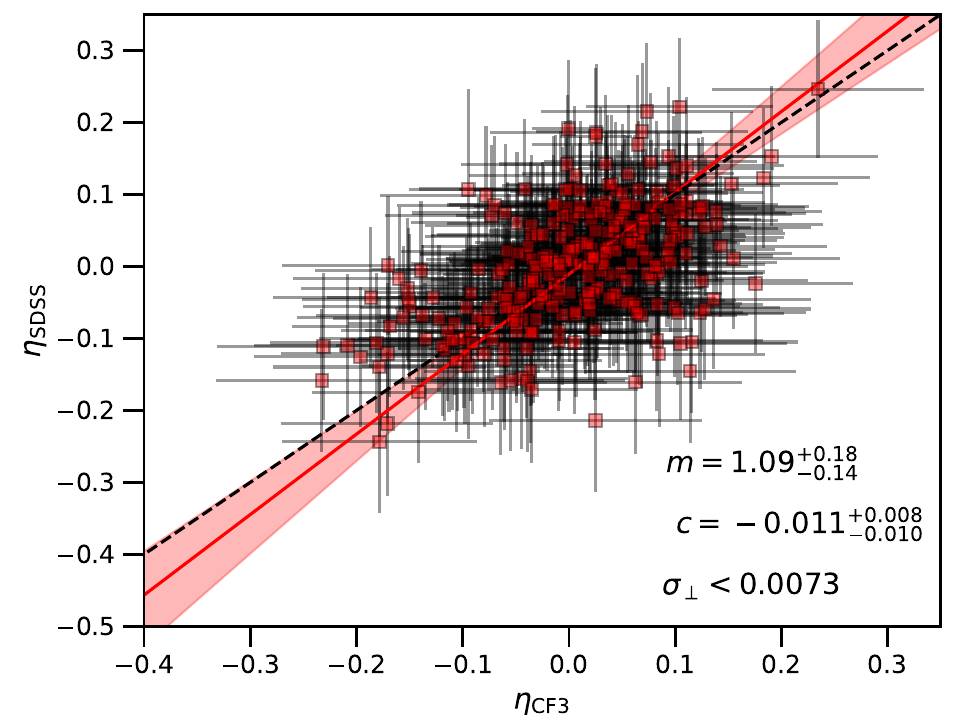}\\
\includegraphics[width=0.48\textwidth, trim=0pt 0pt 0pt 0pt, clip]{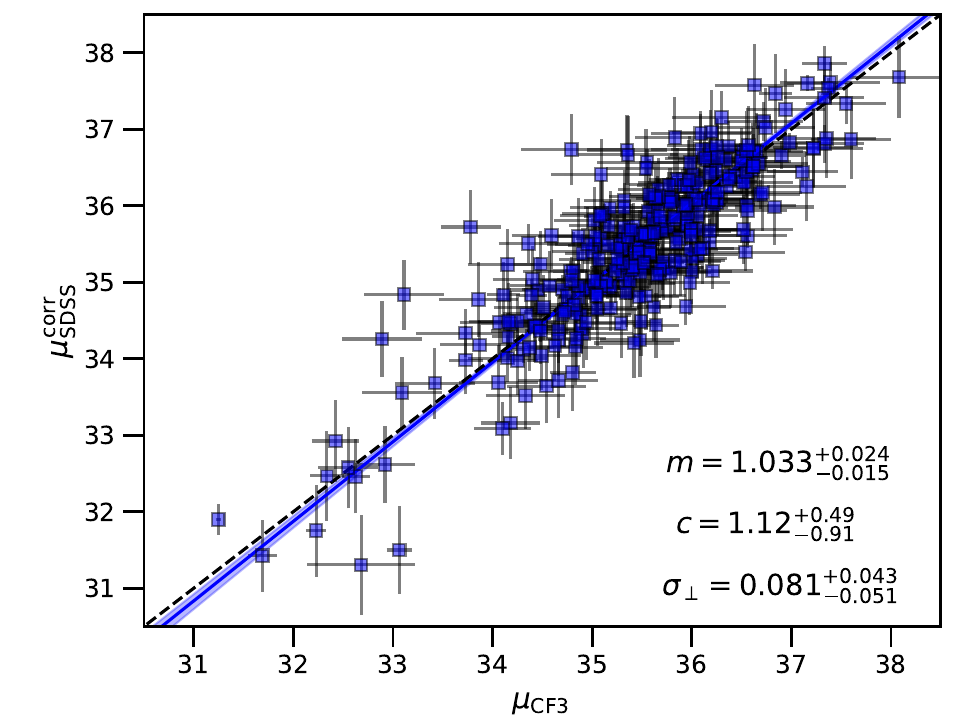}\\
  \caption{A comparison of log-distance ratios from galaxies in common, and distance moduli from groups in common, between the SDSS PV catalogue and CosmicFlows-III. We use log-distance ratios from single FP fits and multiple FP fits for the upper and lower plots respectively. The line and shaded region show the best-fit and corresponding $1\sigma$ region from a fit to the data after the zero-point correction, while the dashed-line is the expected one-to-one line.}
  \label{fig:sdss_zp}
\end{figure}

A comparison of the log-distance ratios for individual objects in both CF3 and the SDSS PV sample, and distance moduli for groups containing overlapping galaxies from both datasets after applying our zero-point correction is shown in Figure~\ref{fig:sdss_zp}. The best-fit and $1\sigma$ shaded regions are obtained taking into account the uncertainties in both axes using \textsc{hyperfit} (Appendix~\ref{app:hyperfit}). The fit after the correction is consistent with a one-to-one line, with small intrinsic scatter, demonstrating excellent agreement between the two independent sets of measurements. Given the close agreement across a wide range of values, we do not apply a change to the slope in addition to the zero-point offset.

\begin{figure}
\centering
\includegraphics[width=0.48\textwidth, trim=0pt 0pt 0pt 0pt, clip]{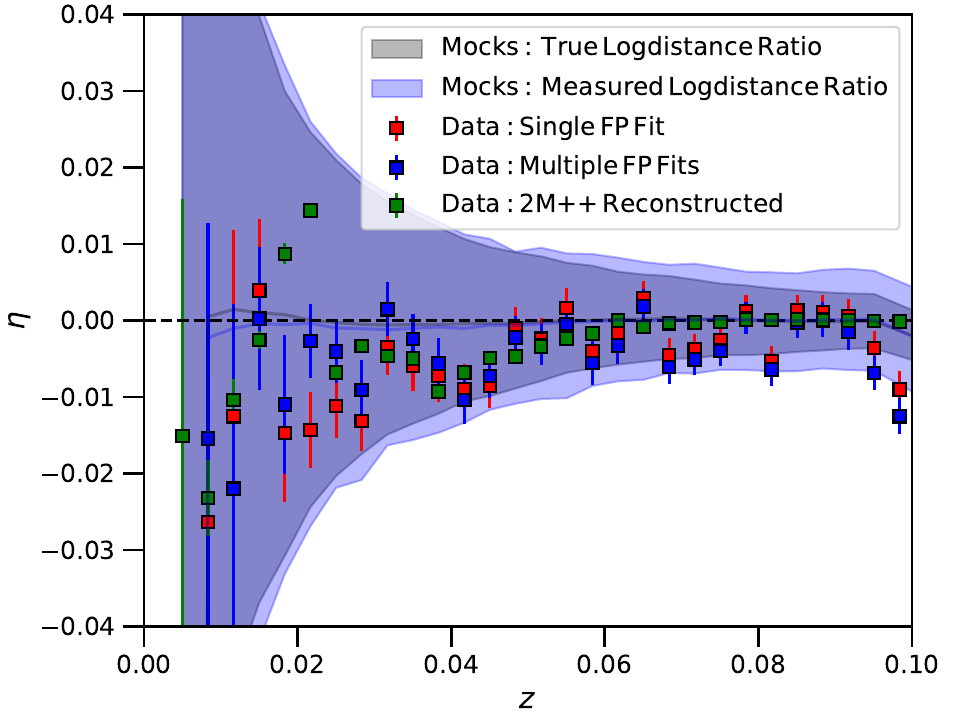}
  \caption{Log-distance ratios in the SDSS PV catalogue (points) and simulations (faint lines/bands), binned as a function of redshift. The grey and blue filled-in bands and corresponding lines show the mean and $95\%$ ($2\sigma$) bounds for the simulations using the true and measured log-distance ratios respectively, which are centred on the horizontal dashed line at $\eta=0$ as expected. We show log-distance ratio measurements from single and multiple FP fits as a function of group richness, and predicted from the 2M++ reconstruction of \protect\cite{Carrick2015}.}
  \label{fig:sdss_hubble}
\end{figure}

This choice is further justified in Figure~\ref{fig:sdss_hubble}, which shows the mean log-distance ratio in bins of redshift. Aside from the presence of large-scale structures along the line-of-sight, there is no evidence of radial systematics in the data (as would be hoped given the mock validation in Section~\ref{sec:red_bias}), and reasonable agreement between the three methods of estimating the log-distance ratio. Note that the reconstruction of \cite{Carrick2015} only extends up to $z=0.067$, so beyond this the predicted velocity is simply forced to gradually tend to zero and does not include any inhomogeneities that may exist at these higher redshifts.

It is important to note that we do not include the uncertainty on $\bar{r}$ of 0.016 (most of which comes from cosmic variance) from Table~\ref{tab:FP} in the zero-point error budget in Eq.~\ref{eq:zeropoint}. As we are comparing multiple measurements for the same objects in this calibration, which are subject to the same cosmic variance, the expected error is hence much smaller than the 0.016 found from our ensemble of mocks. Nonetheless, if one were to compare the log-distance ratios in our catalogue to the prediction from LCDM (wherein cosmic variance must be accounted for), the error in the zero-point that should be considered would be the combination of the zero-point uncertainty from the comparison to CF3 (0.0040), and that due to cosmic variance (0.016). How exactly that is done would depend on the method, as it may be that the cosmic variance contribution is instead incorporated into the theory calculation (as is often done in bulk flow studies, see Section~\ref{sec:cosmo}), and so should be kept separate to avoid double counting.

We also do not include the error in the zero-point from CF3 (which for $H_{0}=75 \pm 2\,\mathrm{km\,s^{-1}\,Mpc^{-1}}$ gives an uncertainty on the log-distance ratio of $0.0116$). We do so as to enable other choices of zero-point, Hubble Constant or distance ladder anchor to be made. Any uncertainties in the calibration of the CF3 distances themselves to Cepheids or other local distance anchors should hence be included as an additional error contribution to the 0.004 we quote.

\subsection{Summary}
In this section, we have provided a thorough explanation for how to extract measurements of the log-distance ratio and peculiar velocity from a sample of Fundamental Plane galaxies. Our approach includes a new analytic method for accounting for selection bias, which makes it tractable for us to apply the same technique to all our mocks. We have demonstrated that the method produces unbiased log-distance ratios, but that a residual correlation between the group richness and log-distance ratio requires us to fit subsamples with different cluster sizes separately. Our final combined catalogue achieves a mean uncertainty on the log-distance ratio of $0.1$ dex, which translates to a $\sim 23\%$ uncertainty on the distance. This is slightly better than was achieved with 6dFGSv ($26\%$; \citealt{Springob2014}), and could potentially be improved further with a more detailed understanding of the observed correlation between group richness and the FP parameters. Finally, we tie the zero-point of our sample to the CosmicFlows-III data, demonstrating consistency using both individual objects and objects within the same cluster, and recovering a relative zero-point uncertainty of $0.004$ dex (not including cosmic variance or the uncertainty in the CF3 zero-point itself).

\section{Bulk Flow}
\label{sec:cosmo}

In the last part of this work, we present measurements of the bulk flow from the data and mocks as an example of the analysis that can be performed using our publicly available SDSS PV catalogue and associated simulations. All measurements are performed in Supergalactic cartesian coordinates.

\begin{figure*}
\centering
\includegraphics[width=\textwidth, trim=0pt 0pt 0pt 0pt, clip]{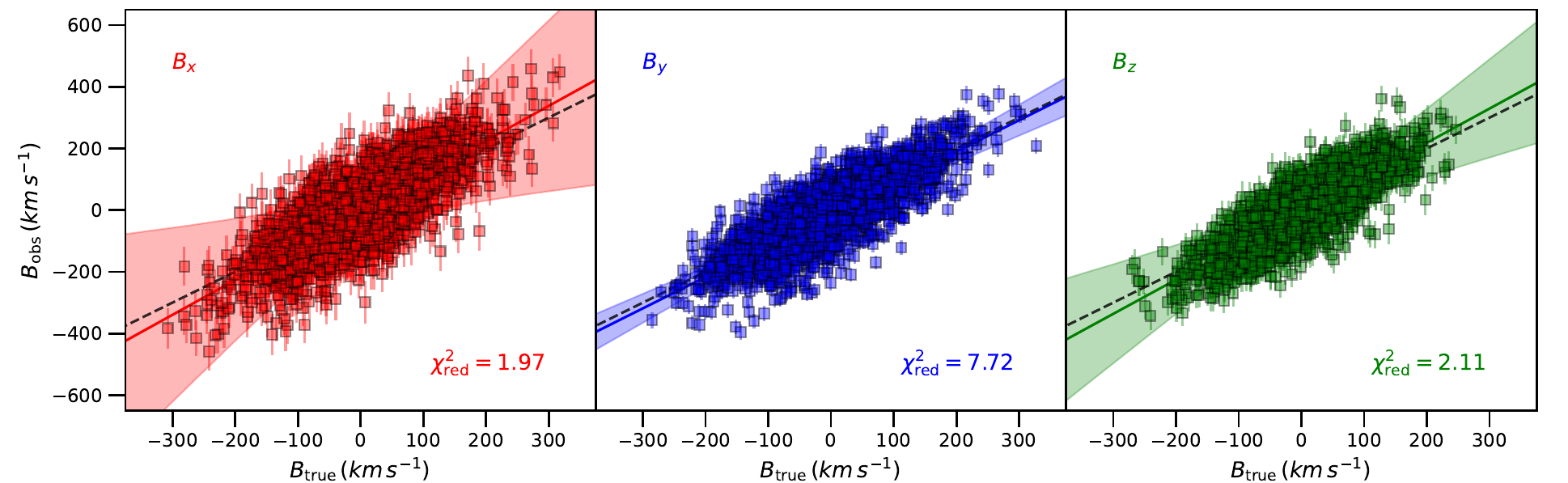}
  \caption{A comparison of the measured and true bulk flow in the Supergalactic $x$, $y$ and $z$ directions (left, middle, and right columns respectively) in the SDSS mocks. The solid lines and shaded regions show a linear fit, plus $1\sigma$ errors on the fit, \textit{where we have assumed the reported errors are accurate}. In this case we recover the expected one-to-one line (dashed black line) to within $1\sigma$. However, the error bars are not fully representative of the scatter seen in the measurements, as can be seen from the reduced chi-squared $\chi^{2}_\mathrm{{red}}$ between the observed and true values from the mocks. This can also be seen in \protect\cite{Qin2018} and \protect\cite{Qin2021b} and thus requires further study in subsequent work to improve.}
  \label{fig:sdssbf}
\end{figure*}

In Figure~\ref{fig:sdssbf}, we show measurements of the bulk flow estimated from our 2048 mock catalogues using the Maximum Likelihood method \citep{Kaiser1988} applied directly to the log-distance ratios as in \cite{Qin2018,Qin2019a}. The `true' bulk flow is defined simply as the weighted average of the underlying velocities in each direction, where the measurement error is used as the weight to ensure that the two sets of bulk flows are at the same effective depth.

Figure~\ref{fig:sdssbf} shows that we recover bulk flow measurements that are, on average, unbiased and well correlated with the true bulk flow in each simulation. However, the error bars do not represent well the scatter between the measured and true values, as can be seen by the very large reduced $\chi^{2}$ difference between the observed and true values from the mocks in each of the three separate directions included in the figure. A similar result can be seen in \cite{Qin2018}, \cite{Qin2021b} and other work using the Maximum Likelihood Estimator. Possible reasons for this are a failure of our assumption that the velocity of each galaxy can be represented simply as a bulk flow, without higher order components, and/or that the distribution of each measured velocity can be treated as an independent Gaussian. However, we leave detailed testing of this hypothesis, and of whether the scatter can be reduced or the error bars made more reasonable, for future work. Instead, when necessary in this work, we simply enlarge the observational errors in each of the three components by a factor equal to $\sqrt{\chi^{2}_{\mathrm{red}}}$, such that the reduced chi-squared is renormalised to one.\footnote{The reduced $\chi^{2}$ in Figure~\ref{fig:sdssbf} has been computed assuming the three directions are independent. This is not true in practice as they are fit at the same time from data with only radial peculiar velocities. However, we verified that $\chi^{2}_{\mathrm{red}}$ is similar when using the full $3\times3$ covariance for each mock or just summing the individual $\chi^{2}$ values for each direction. This is demonstrated further by the fact that after rescaling the uncertainty on each \textit{individual} direction so that the reduced chi-squared for each component is one, the reduced chi-squared accounting for the cross-correlation is also close to one ($\chi^{2}_{\mathrm{red}} = 1.08$).}

We also find that the uncertainty is smallest and most underestimated in the $y$ axis. This is interesting because our use of the Supergalactic coordinate system places the $y$ axis almost perfectly along the observer's line of sight, with $x$ and $z$ transverse to this. The SDSS PV survey is a somewhat narrow but long cone (compared to a survey like 6dFGSv) and so the large $\chi^{2}_{\mathrm{red}}$ in this direction indicates that there is likely some systematic in the measurement technique (i.e., neglecting higher-order moments) that becomes more important when we are able to average over a larger volume.

\begin{table}
\centering
\caption{Estimates of the bulk flow from the SDSS data. `Data' columns correspond to bulk flows measured from the SDSS PV catalogue using either a single FP fit to the entire sample, or our preferred method of fitting separate FP's to the sample as a function of group richness to remove the bias demonstrated in Section~\ref{sec:groupcorr}. In both cases, the uncertainties have been increased by a constant factor to correct for the underestimation of the observational uncertainties seen in the mocks (Fig.~\ref{fig:sdssbf}). `$\Lambda$CDM' columns correspond to the predictions from our fiducial cosmological model for the SDSS survey geometry. We list the three individual components $B_{i}$ (for which the expected value is always 0), the bulk flow amplitude $|\boldsymbol{B}|$, and the weighted depth of the measurements $d_{\mathrm{MLE}}$. The last row gives the probability of recovering a $\chi^{2}$ difference between the data and $\Lambda$CDM that is \textit{larger} than the value we actually find. For the individual components the expectation is a Gaussian with zero-mean and standard deviation $\sim$120\,km\,s$^{-1}$ and for the bulk flow amplitude (which is Maxwell-Boltzmann distributed) we list the `most probable' value and $68\%$ confidence limits.}
\begin{tabular}{lcccc} \hline
 & \multicolumn{2}{c}{Single FP fit} & \multicolumn{2}{c}{Multiple FP fits} \\
 & Data & $\Lambda$CDM & Data & $\Lambda$CDM  \\ \hline \vspace{1pt}
$B_{x}\,(\mathrm{km\,s^{-1}})$ & $-224^{+73}_{-97}$ & $\pm118$ & $-345^{+66}_{-101}$ & $\pm119$ \vspace{2pt} \\
$B_{y}\,(\mathrm{km\,s^{-1}})$ & $-106^{+39}_{-34}$ & $\pm131$ & $-88^{+40}_{-29}$ & $\pm131$ \vspace{2pt} \\
$B_{z}\,(\mathrm{km\,s^{-1}})$ & $~190^{+140}_{-105}$ & $\pm107$ & $-75^{+107}_{-132}$ & $\pm107$ \vspace{2pt} \\
$|\boldsymbol{B}|\,(\mathrm{km\,s^{-1}})$ & $~323^{+82}_{-76}$ & $98^{+50}_{-43}$ & $381^{+85}_{-79}$ & $99^{+51}_{-43}$ \vspace{4pt} \\
$d_{\mathrm{MLE}}\,(h^{-1}\mathrm{Mpc})$ & \multicolumn{2}{c}{139} & \multicolumn{2}{c}{139} \vspace{2pt} \\
$P(>\chi^{2})$ & \multicolumn{2}{c}{20.0$\%$} & \multicolumn{2}{c}{10.8$\%$}
\vspace{1pt} \\ \hline
\end{tabular}
\label{tab:sdssbf}
\end{table}

\begin{figure*}
\centering
\includegraphics[width=\textwidth, trim=0pt 0pt 0pt 0pt, clip]{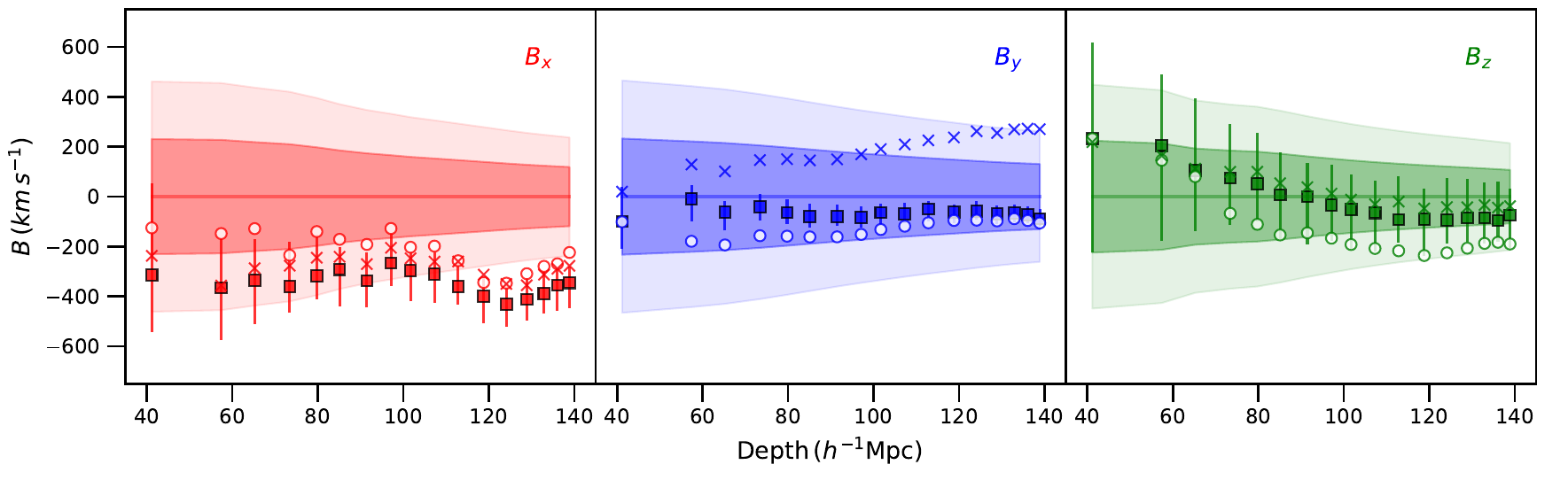}
  \caption{Bulk flow measurements from the SDSS PV catalogue as a function of weighted depth. Each point is computed by cutting the full catalogue to different values of $z_{\mathrm{max}} \in (0.02, 0.10)$ with $\Delta z_{\mathrm{max}}=0.005$. Circles and squares show measurements using log-distance ratios for the sample with single and multiple FP fits as a function of group richness respectively. Crosses are measurements for the multiple FP-fit sample, but where we offset the zero-point by $+0.01$ dex (a $2.5\sigma$ shift compared to our zero-point uncertainty), to demonstrate that such errors have only a small affect on the $x$ and $z$ directions. Error bars have been corrected for the underestimation seen in the mocks, and are of similar size for all three data samples so are only plotted for one set for clarity.  The shaded regions show the $1$ and $2\sigma$ expectations for our fiducial $\Lambda$CDM model.}
  \label{fig:sdssbfdata}
\end{figure*}

Our results applying the same procedure to the full set of data are shown in Table~\ref{tab:sdssbf}. In Fig.~\ref{fig:sdssbfdata}, we show the same measurements but cutting the upper redshift limit of the data at seventeen different values of $z_{\mathrm{max}} \in (0.02, 0.10)$ with $\Delta z_{\mathrm{max}}=0.005$. When making such measurements, we remove any data with log-distance ratios scattered more than $4\sigma$ from the mean to ensure objects with outlying peculiar velocities do not bias our results. We also correct all the observational uncertainties for the data by the same factor as was found to be necessary to bring the true and measured values for the mocks into statistical agreement. We compute this scaling separately for each value of $z_{\mathrm{max}}$. 

In both the table and figure we also provide the theoretical, cosmic-variance expectation of a $\Lambda$CDM model with our fiducial/simulation cosmology (given at the end of Section~\ref{sec:introduction}) accounting for the actual geometry of the SDSS PV catalogue and the uncertainty on each measurement. This is done using the method of \cite{Feldman2010} and \cite{Ma2011}, where the theoretical covariance between each peculiar velocity measurement is computed using
\begin{equation}
    \mathcal{W}_{mn} = \frac{\Omega_{m}^{1.1}H^{2}_{0}}{2\pi^{2}}\int f_{mn}(k)P(k)dk ~,
\end{equation}
which depends on both the fiducial cosmological model (through the matter power spectrum, $P(k)$, matter density $\Omega_{m}$ and Hubble constant $H_{0}$) and the relative location of each pair of galaxies in the SDSS PV sample (through $f_{mn}$; which is given in Equation A11 of \citealt{Ma2011}). For the SDSS PV catalogue $\mathcal{W}_{mn}$ is a $34,059\times34,059$ matrix. To reduce this down to the theoretical covariance matrix for the 3 bulk flow components $R^{(v)}_{pq}$, we multiply by the vector of Maximum Likelihood weights for each galaxy $n$, $R_{pq} = w_{p,m}w_{q,n}\mathcal{W}_{mn}$ (where we adopt the Einstein summation convention), and where
\begin{equation}
    w_{p,n} = A^{-1}_{pj}\frac{\hat{\boldsymbol{x}}_{j,n}}{\sigma^{2}_{v,n} + \sigma^{2}_{*}} ~; \quad A_{pj} = \frac{\hat{\boldsymbol{x}}_{p,n}\hat{\boldsymbol{x}}_{j,n}}{\sigma^{2}_{v,n} + \sigma^{2}_{*}} ~.
\end{equation}
These weights depend only on the unit vector defining the position of the galaxy with respect to our 3 bulk flow directions $\hat{\boldsymbol{x}}$, and the uncertainty on the peculiar velocity measurement. We convert the errors on the log-distance ratio to those on velocities using the approximation of \cite{Watkins2015}, $\sigma_{v,n} = \ln(10) cz_{\mathrm{mod}}/(1+z_{\mathrm{mod}})\,\times\,\sigma_{\eta,n}$, and add a small additional contribution to the measurement uncertainties, $\sigma_{*} = 350\,\mathrm{km\,s^{-1}}$ to account for non-linearities in the velocity field. In this way, the weights encode the relative contribution of each galaxy in the sample to the overall bulk flow measurement, and so the theoretical prediction also takes into account that galaxies with larger errors will contribute less, and galaxies with radial peculiar velocities aligned with, for example, the $y$-direction will not contribute to the $x$ or $z$ direction bulk flow.

Finally, we compute the chi-squared value for a measured bulk flow $\boldsymbol{B}$ given our theoretical model using 
\begin{equation}
    \chi^{2}_{BF} = B_{p}(R^{(v)}_{pq} + \beta_{p}\beta_{q}R^{(\epsilon)}_{pq})^{-1}B^{T}_{q}
\end{equation}
where we have included the measurement error for the bulk flow, $R^{(\epsilon)}_{pq}$ scaled by $\boldsymbol{\beta}$ to account for the underestimation of the errors seen in the mocks (Fig~\ref{fig:sdssbf}).

Considering both cosmic variance and the measurement error, the probability of obtaining a larger bulk flow in $\Lambda$CDM is $20.0\%$ and $10.8\%$ for the single and multiple FP fit samples respectively---too high to rule out the null hypothesis (that our fiducial $\Lambda$CDM model is correct). However, looking in more detail at the measurements when cutting the SDSS data at different depths, we see that there is a preference for a larger-than-expected bulk flow in the $x$ direction that is persistent when including data above $z=0.08$ and actually more discrepant with the $\Lambda$CDM prediction when the highest redshift data is \textit{not} included. Using only data for $z\leq0.08$, we find $P(>\!\!\chi^{2}) = 8.1\%$ and $6.1\%$ for the single and multiple FP fits respectively, which is closer to the $5\%$ confidence level but still not enough to disfavour $\Lambda$CDM with confidence. The measurements using our preferred multiple FP fits as a function of group richness are not quite consistent with those for a single FP fit to the full sample, which is perhaps not surprising given the bias in the `single FP' sample identified in Section~\ref{sec:groupcorr}. Nonetheless, we present both to highlight that (regardless of the choice of data) both recover slightly larger than expected bulk flows at large distances, at least in the $x$-direction.

This is interesting because several other studies have reported larger-than-expected bulk flows at similar depth \citep{Pike2005,Feldman2008,Kashlinsky2008,Feldman2010,Lavaux2010}. One possible explanation for our result is systematic errors, but we would typically expect the impact of any residual systematics to only become larger when the higher redshift data is included and, moreover, the choice of coordinates used in this bulk flow analysis places the axis of increasing redshift (which is the one most prone to systematic errors) almost purely in the Cartesian $y$-direction. It is in this direction that any zero-point calibration errors would be mostly confined --- as can be seen in Fig~\ref{fig:sdssbfdata}, a change to the zero-point of $+0.01$ dex (a $2.5\sigma$ change) causes only a small change in the $x$ and $z$-direction bulk flow compared to the $y$-direction. 

\begin{figure}
\centering
\includegraphics[width=0.48\textwidth, trim=5pt 40pt 10pt 30pt, clip]{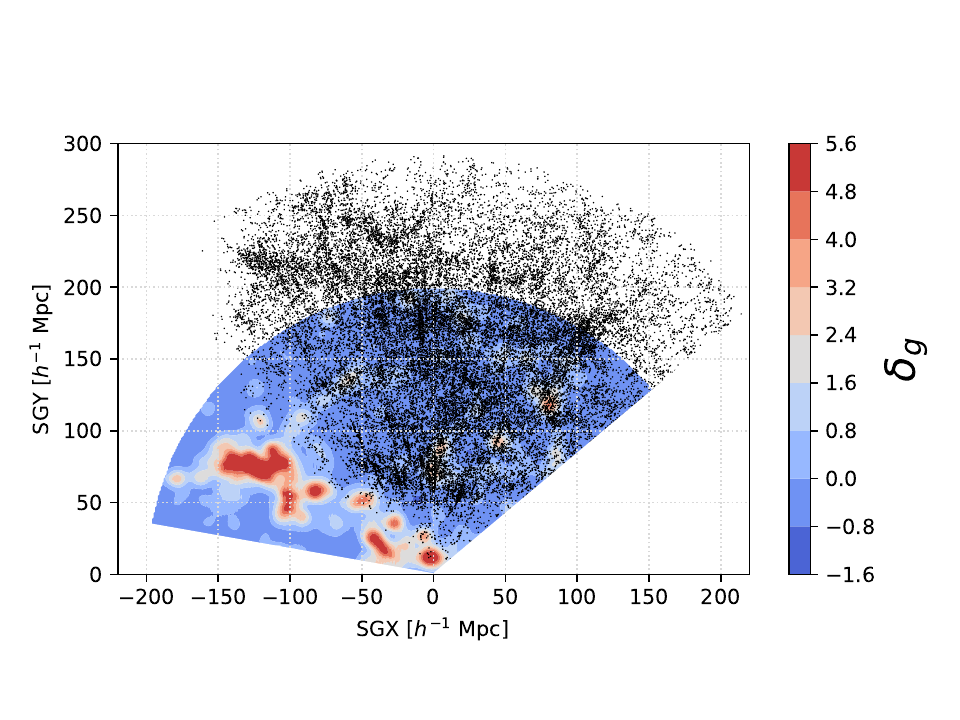}
  \caption{Map of the SDSS PV catalogue (black points) alongside a $-1\,h^{-1}\mathrm{Mpc} < \mathrm{SGZ} < 1\,h^{-1}\mathrm{Mpc}$ slice of the 2M++ reconstruction of the local density field \protect\citep{Carrick2015} in Supergalactic coordinates. The large overdensity at $\mathrm{SGX}\approx -125\,h^{-1}\mathrm{Mpc}$, $\mathrm{SGY}\approx 75\,h^{-1}\mathrm{Mpc}$ is the Shapley supercluster, which provides a possible explanation for our larger-than-expected bulk flow measurement.}
  \label{fig:footprint}
\end{figure}

As a physical explanation, it is worth noting that the direction of our measured bulk flow aligns well with the position of the Shapley supercluster, as can be seen in Fig.~\ref{fig:footprint} using the reconstructed density field from 2M++ \citep{Carrick2015}. However, it is difficult to say whether the amplitude of our measurement is in agreement with what would be expected given the gravitational influence from this structure. This hence provides an interesting avenue for further study, either with improved techniques for measuring the bulk flow (such as the Minimum Variance estimator; \citealt{Watkins2009}) such that we can avoid having to correct the observational uncertainties as done here, or with data at the same or larger depths and over a wider area, which may be possible with upcoming surveys.

\section{Conclusions}

In this paper we present the Sloan Digital Sky Survey Peculiar Velocity catalogue, a collection of $34,059$ high-quality peculiar velocity measurements up to $z=0.1$, subtending an area of $7016\,\mathrm{deg}^{2}$. We also provide a detailed analysis of the characteristics of the data, identifying and correcting systematic errors, and find excellent agreement with overlapping measurements from existing surveys. A key finding is the trend between mean surface brightness, slope $b$, and group richness. Whether this reflects an intrinsic dependence of the FP due to the differing assembly histories in different environments, or is due to unresolved systematics remains unclear. However, for the purposes of the PV catalogue, we account for this by fitting separate FPs to samples of different group richness. 

Alongside our data, we make public an ensemble of 2,048 simulated catalogues that almost exactly reproduce the selection function and quality of the data and were run through the same pipeline to enable accurate systematic calibration. As a necessary step towards this, we created improved techniques for fast fitting of the FP and extraction of peculiar velocities, which also set the stage for next-generation samples of data that we may expect from upcoming surveys such as DESI \citep{DESI2016} or southern hemisphere surveys on the 4MOST facility (4HS; PIs Cluver and Taylor). However, we caution that the 3D Gaussian model that we assume here may need to be extended to better incorporate skewness arising from the inclusion of fainter galaxies from these samples --- in which case our analytic corrections for the  selection functions will also need to be revisited.  

In terms of future work, our preliminary tests weakly suggest a bulk flow from the SDSS PV data that is higher than expected from $\Lambda$CDM. We have demonstrated that the alignment of this flow implies it is not a result of either our correction for group richness or an inaccurate zero-point calibration, but could be a result of the proximity of our data to the Shapley supercluster. Further work is required to uncover the full origin of this large bulk flow. We hope that the publicly available SDSS PV data and associated data products will provide all the necessary ingredients to do just that, as well as enabling further cosmological and cosmographic analysis of our local Universe.

\section*{Acknowledgements}

CH and KS acknowledge support from the Australian Government through the Australian Research Council’s Laureate Fellowship funding scheme (project FL180100168). 
JRL acknowledges support from the UK Science and Technology Facilities Council through the Durham Astronomy Consolidated Grants ST/P000541/1 and ST/T000244/1.
FQ is supported by the project \begin{CJK}{UTF8}{mj}우주거대구조를 이용한 암흑우주 연구\end{CJK} (``Understanding Dark Universe Using Large Scale Structure of the Universe''), funded by the Ministry of Science. MC acknowledges support from the Australian Government through the Australian Research Council’s Discovery Projects funding scheme (project DP160102075). This research has made use of NASA's Astrophysics Data System Bibliographic Services and the \texttt{astro-ph} pre-print archive at \url{https://arxiv.org/}, the {\sc matplotlib} plotting library \citep{Hunter2007}, and the {\sc chainconsumer} and {\sc emcee} packages \citep{Hinton2016, ForemanMackey2013}. Computations were performed on the OzSTAR national facility at Swinburne University of Technology, which receives funding in part from the Astronomy National Collaborative Research Infrastructure Strategy (NCRIS) allocation provided by the Australian Government, and with the assistance of resources and services from the National Computational Infrastructure (NCI), which is also supported by the Australian Government.

\section*{Data Availability}

The SDSS PV catalogue and associated data products and simulations are available on Zenodo: \url{https://zenodo.org/record/6640513}. The catalogue can also be accessed in a modified form with slight additional metadata at the Extragalactic Distance Database \url{\eddurl} in the section Summary Distances, in a file called FP: SDSS Distances. Raw SDSS data was obtained from the \href{https://skyserver.sdss.org/casjobs/}{SDSS Casjobs} server. Exact queries used for the SDSS PV data and its supersets in Table~\ref{tab:selectioncriteria} will be shared upon reasonable request to the corresponding author, as will all other codes or data.


\bibliographystyle{mnras}
\bibliography{massive.bib}


\appendix

\section{HyperFit}
\label{app:hyperfit}

In the course of this work, we have frequently required a fast, simple method to fit a line or plane to data allowing for either, or both, intrinsic scatter and (potentially correlated) errors on all the input variables (i.e.\ in both the `x' and 'y' variables for a 2D fit). A general method for such fitting is detailed in \cite{Robotham2015} and implemented through the associated R-package \textsc{hyperfit}. For our purposes we found it useful to produce a similar package in Python. This package has been fully documented and made `pip-installable'. It provides vectorised methods to simply find the best fit given the data or to return a full set of posterior samples for the model given the data. Real, astrophysical, test data is provided with the package, demonstrating that it typically takes only a few seconds to find the best fit or a couple of minutes for a fully converged MCMC run. More details can be found at \url{https://hyperfit.readthedocs.io/en/latest/}.

\section{Transformation between FP eigenvectors and parameters}
\label{app:FPvectors}

In the most general 3D Gaussian method \citep{Saglia2001,Colless2001,Magoulas2012}, the FP is defined by three orthonormal unit eigenvectors,
\begin{align}
\boldsymbol{\hat{v}}_{1} &= \frac{\boldsymbol{\hat{r}}-a\boldsymbol{\hat{s}}-b\boldsymbol{\hat{i}}}{|\boldsymbol{v_{1}}|} \notag \\
\boldsymbol{\hat{v}}_{2} &= \frac{b\boldsymbol{\hat{r}}-bk\boldsymbol{\hat{s}}+(1-ka)\boldsymbol{\hat{i}}}{|\boldsymbol{v_{2}}|} \notag \\
\boldsymbol{\hat{v}}_{3} &= \frac{(ka^{2}-a\boldsymbol+kb^{2}){\hat{r}}+(ka-1-b^{2})\boldsymbol{\hat{s}}+(kb+ab)\boldsymbol{\hat{i}}}{|\boldsymbol{v_{1}}||\boldsymbol{v_{2}}|}
\end{align}
where 
\begin{align}
|\boldsymbol{v_{1}}| &= \sqrt{(1+a^{2}+b^{2})} \notag \\ 
|\boldsymbol{v_{2}}| &= \sqrt{1+b^{2}+k^{2}(a^{2}+b^{2}-2a/k)} ~.
\end{align}
Using these expressions and the Jacobian we can write the scatter matrix components for the FP parameter space shown in Eq.~\ref{eq:covariance} in terms of $\sigma_{1}$, $\sigma_{2}$, $\sigma_{3}$ (the scatter in each of the orthonormal coordinates) as
\begin{align}
    \sigma^{2}_{r} &= \frac{\sigma^{2}_{1}}{|\boldsymbol{v_{1}}|^{2}} + \frac{b^{2}\sigma^{2}_{2}}{|\boldsymbol{v_{2}}|^{2}} + \frac{(ka^{2}-a+kb^{2})^{2}\sigma^{2}_{3}}{|\boldsymbol{v_{1}}|^{2}|\boldsymbol{v_{2}}|^{2}} \\
    \sigma^{2}_{s} &= \frac{a^{2}\sigma^{2}_{1}}{|\boldsymbol{v_{1}}|^{2}} + \frac{k^{2}b^{2}\sigma^{2}_{2}}{|\boldsymbol{v_{2}}|^{2}} + \frac{(ka-1-b^{2})^{2}\sigma^{2}_{3}}{|\boldsymbol{v_{1}}|^{2}|\boldsymbol{v_{2}}|^{2}} \\
    \sigma^{2}_{i} &= \frac{b^{2}\sigma^{2}_{1}}{|\boldsymbol{v_{1}}|^{2}} + \frac{(1-ka)^{2}\sigma^{2}_{2}}{|\boldsymbol{v_{2}}|^{2}} + \frac{(kb+ab)^{2}\sigma^{2}_{3}}{|\boldsymbol{v_{1}}|^{2}|\boldsymbol{v_{2}}|^{2}} \\
    \sigma_{rs} &= -\frac{a\sigma^{2}_{1}}{|\boldsymbol{v_{1}}|^{2}} - \frac{kb^{2}\sigma^{2}_{2}}{|\boldsymbol{v_{2}}|^{2}} + \frac{(ka-a+kb^{2})(ka-1-b^{2})\sigma^{2}_{3}}{|\boldsymbol{v_{1}}|^{2}|\boldsymbol{v_{2}}|^{2}} \\
    \sigma_{ri} &= -\frac{b\sigma^{2}_{1}}{|\boldsymbol{v_{1}}|^{2}} + \frac{b(1-ka)\sigma^{2}_{2}}{|\boldsymbol{v_{2}}|^{2}} + \frac{(ka-a+kb^{2})(kb+ab)\sigma^{2}_{3}}{|\boldsymbol{v_{1}}|^{2}|\boldsymbol{v_{2}}|^{2}} \\
    \sigma_{si} &= \frac{ab\sigma^{2}_{1}}{|\boldsymbol{v_{1}}|^{2}} - \frac{kb(1-ka)\sigma^{2}_{2}}{|\boldsymbol{v_{2}}|^{2}} + \frac{(ka-1-b^{2})(kb+ab)\sigma^{2}_{3}}{|\boldsymbol{v_{1}}|^{2}|\boldsymbol{v_{2}}|^{2}} ~.
\end{align}

In practice, previous works \citep{Saglia2001,Colless2001,Magoulas2012} all found that the second eigenvector has a very weak dependence on $s$, so that the longest axis of the 3D Gaussian is confined to the $r$--$i$ plane. If one assumes \textit{a priori} that this is true, then $k=0$ in the above expressions. We make the same assumption in our fits.

\section{Derivation of Analytic \texorpdfstring{$\mathbf{f_{n}}$}{}}
\label{sec:appfn1}

In this appendix we derive the expression for the integral over the 3D Gaussian of the FP in terms of elementary functions. Although the exact derivation here is somewhat specific to the typical selection function imposed on FP measurements, it can be adapted to other scenarios requiring the integral over a 3D Gaussian function.\footnote{This derivation is heavily indebted to anonymous user \textsc{Przemo}'s derivation of the \href{https://math.stackexchange.com/questions/869502/multivariate-gaussian-integral-over-positive-reals/}{`Multivariate Gaussian integral over positive reals'} on \textsc{Mathematics Stack Exchange}.}

\subsection{General case}

The integral we seek to solve has the form
\begin{multline}
    f_{n} = \frac{1}{(2\pi)^{3/2}|\boldsymbol{\mathsf{C}}|^{1/2}}\int_{-\infty}^{\infty}\int_{r_{\mathrm{min}}-i/2}^{r_{\mathrm{max}}-i/2}\int_{s_{\mathrm{min}}}^{\infty}ds dr di \\ \exp\biggl\{-\frac{1}{2}(\boldsymbol{x}-\boldsymbol{\bar{x}})\boldsymbol{\mathsf{C}}^{-1}(\boldsymbol{x}-\boldsymbol{\bar{x}})^{T}\biggl\} ~,
\end{multline} 
where $\boldsymbol{x} = (r, s, i)$, $\boldsymbol{\bar{x}} = (\bar{r},\bar{s},\bar{i})$ and
\begin{equation}
    \boldsymbol{\mathsf{C}}^{-1} = 
    \begin{pmatrix}
    \Psi_{rr} & \Psi_{rs}  & \Psi_{ri} \\
    \Psi_{rs} & \Psi_{ss}  & \Psi_{si} \\
    \Psi_{ri} & \Psi_{si}  & \Psi_{ii}
    \end{pmatrix} ~.
\end{equation}
For our FP scenario, the values of $r_{\mathrm{max(min)}}$ and the covariance matrix $\boldsymbol{\mathsf{C}}$ change for each galaxy $n$; $r_{\mathrm{max(min)}}$ also vary as a function of distance. The first thing to note is that we are free to re-centre the integral about the mean values by performing a simple change of base,
\begin{multline}
    f_{n} = \frac{1}{(2\pi)^{3/2}|\boldsymbol{\mathsf{C}}|^{1/2}}\int_{-\infty}^{\infty}\int_{r_{\mathrm{min}}-\bar{r}-\bar{i}/2-i/2}^{r_{\mathrm{max}}-\bar{r}-\bar{i}/2-i/2}\int_{s_{\mathrm{min}}-\bar{s}}^{\infty}ds dr di \\ \exp\biggl\{-\frac{1}{2}\boldsymbol{x}\boldsymbol{\mathsf{C}}^{-1}\boldsymbol{x}^{T}\biggl\} ~.
    \label{eq:app1}
\end{multline} 
Focussing on the integrand, we can expand the exponent as
\begin{multline}
    \boldsymbol{x}\boldsymbol{\mathsf{C}}^{-1}\boldsymbol{x}^{T} = \Psi_{ss}\biggl(s+\frac{\Psi_{si}i+\Psi_{rs}r}{\Psi_{ss}}\biggl)^{2} + \biggl(\Psi_{ii}-\frac{\Psi_{si}^{2}}{\Psi_{ss}}\biggl)i^{2} + \\ 
    \biggl(\Psi_{rr}-\frac{\Psi_{rs}^{2}}{\Psi_{ss}}\biggl)r^{2} + 2\biggl(\Psi_{ri}-\frac{\Psi_{rs}\Psi_{si}}{\Psi_{ss}}\biggl)ri ~.
\end{multline}
Substituting this into Eq. ~\ref{eq:app1}, we can solve the integral over $s$,
\begin{align}
    f_{n} &= \frac{1}{4\pi \sqrt{\Psi_{ss}|\boldsymbol{\mathsf{C}}|}}\int_{-\infty}^{\infty}\int_{r_{\mathrm{min}}-\bar{r}-\bar{i}/2-i/2}^{r_{\mathrm{max}}-\bar{r}-\bar{i}/2-i/2} dr di \, \exp\biggl\{-\frac{1}{2} \times \notag \\
    & \biggl[\biggl(\Psi_{ii}-\frac{\Psi_{si}^{2}}{\Psi_{ss}}\biggl)i^{2} + \biggl(\Psi_{rr}-\frac{\Psi_{rs}^{2}}{\Psi_{ss}}\biggl)r^{2} + 2\biggl(\Psi_{ri}-\frac{\Psi_{rs}\Psi_{si}}{\Psi_{ss}}\biggl)r i\biggl]\biggl\} \notag \\
    & \mathrm{erfc}\biggl(\frac{\Psi_{si}i+\Psi_{rs}r+\Psi_{ss}(s_{\mathrm{min}}-\bar{s})}{\sqrt{2\Psi_{ss}}}\biggl) ~,
\end{align}
where $\mathrm{erfc}(u) = 1 - \mathrm{erf}(u)$ is the complementary error function. We now make the substitution $u = (\Psi_{si}i+\Psi_{rs}r+\Psi_{ss}(s_{\mathrm{min}}-\bar{s}))/\sqrt{2\Psi_{ss}}$, so that 
\begin{align}
f_{n} &= \frac{-\exp\biggl\{-\frac{1}{2}(s_{\mathrm{min}}-\bar{s})^{2}\Psi_{ss}\biggl(\frac{\Psi_{ss}\Psi_{ii}}{\Psi_{si}^{2}}-1\biggl)\biggl\}}{\Psi_{si}\sqrt{8\pi^{2}|\boldsymbol{\mathsf{C}}|}}\int_{-\infty}^{\infty}\int_{\ell_{\mathrm{min}}}^{\ell_{\mathrm{max}}} drdu \notag \\
& \mathrm{erfc}(u)\exp\biggl\{-\biggl(\frac{\Psi_{ss}\Psi_{ii}}{\Psi_{si}^{2}}-1\biggl)\biggl(u^{2}-\sqrt{2\Psi_{ss}}(s_{\mathrm{min}}-\bar{s})u\biggl)\biggl\} \notag \\
& \exp\biggl\{-\frac{1}{2}\biggl[\frac{2\sqrt{2\Psi_{ss}}}{\Psi_{si}}\biggl(\Psi_{ri}-\frac{\Psi_{rs}\Psi_{ii}}{\Psi_{si}}\biggl)\biggl(u-\sqrt{2\Psi_{ss}}(s_{\mathrm{min}}-\bar{s})\biggl)r \notag \\
& \qquad \qquad +\frac{\Psi_{rs}}{\Psi_{si}}\biggl(\frac{\Psi_{rs}\Psi_{ii}}{\Psi_{si}}+\frac{\Psi_{rr}\Psi_{si}}{\Psi_{rs}}-2\Psi_{ri}\biggl)r^{2}\biggl]\biggl\} ~,
\end{align}
where
\begin{align}
    \ell_{\mathrm{min(max)}}=\frac{2\Psi_{si}(r_{\mathrm{min(max)}}-\bar{r}-\bar{i}/2)-\sqrt{2\Psi_{ss}}u+\Psi_{ss}(s_{\mathrm{min}}-\bar{s})}{2\Psi_{si}-\Psi_{rs}}.
\end{align}
At first glance it may seem that our choice of substitution is a poor one and the resulting expression is untenable. However, what we have actually done is isolate the parts of the integral that depend on $r$ in a single exponential. In doing so, we have arrived at an expression of the form $\int_{\ell_{\mathrm{min}}}^{\ell_{\mathrm{min}}} \exp[-1/2(Ar+Br{^2})] dr$, which can be expressed as a difference of error functions times an exponential. After performing this integral, substituting $\ell_\mathrm{min(max)}$, and an exhaustive amount of algebra,
\begin{align}
f_{n} &= \frac{1}{4\sqrt{\pi\delta|\boldsymbol{\mathsf{C}}|}}\int_{-\infty}^{\infty} du\, \mathrm{erfc}(u)\exp\biggl\{-\frac{1}{\delta|\boldsymbol{\mathsf{C}}|}f^{2}(u)\biggl\} \notag \\
& \biggl[\mathrm{erf}\biggl\{\frac{Gf(u)}{\sqrt{\delta}}+R_{\mathrm{min}}\biggl\}-\mathrm{erf}\biggl\{\frac{Gf(u)}{\sqrt{\delta}}+R_{\mathrm{max}}\biggl\}\biggl] ~,
\end{align}
where 
\begin{align}
    f(u) &= u-\sqrt{\frac{\Psi_{ss}}{2}}(s_{\mathrm{min}}-\bar{s}), \\
    R_{\mathrm{min(max)}} &=\frac{\sqrt{2\delta}(r_{\mathrm{min(max)}}-\bar{r}-\bar{i}/2)}{2\Psi_{si}-\Psi_{rs}} ~, \label{eq:appdef1}\\
    \delta &= \Psi_{rr}\Psi_{si}^{2} + \Psi_{ii}\Psi_{rs}^{2}-2\Psi_{si}\Psi_{rs}\Psi_{ri} ~, \\
    G & = \sqrt{\Psi_{ss}}\frac{\Psi_{ri}(2\Psi_{si}+\Psi_{rs})-\Psi_{rr}\Psi_{si}-2\Psi_{ii}\Psi_{rs}}{2\Psi_{si}-\Psi_{rs}} ~. \label{eq:appdef2}
\end{align}
We have managed to reduce the original 3D Gaussian integral to a single integration. Although complex to write, this is significantly faster to compute. However, we can go further still. To make clear how, we rewrite this expression using the substitution $x = \sqrt{\frac{1}{\delta|\boldsymbol{\mathsf{C}}|}}f(u)$, so
\begin{align}
f_{n} &= \frac{1}{4\sqrt{\pi}}\int_{-\infty}^{\infty} dx\, \exp\{-x^{2}\}\mathrm{erfc}\biggl\{\sqrt{\delta|\boldsymbol{\mathsf{C}}|}x + \sqrt{\frac{\Psi_{ss}}{2}}(s_{\mathrm{min}}-\bar{s})\biggl\} \notag \\
& \quad \biggl[\mathrm{erf}\biggl\{G\sqrt{|\boldsymbol{\mathsf{C}}|}x+R_{\mathrm{min}}\biggl\}-\mathrm{erf}\biggl\{G\sqrt{|\boldsymbol{\mathsf{C}}|}x+R_{\mathrm{max}}\biggl\}\biggl] ~.
\label{eq:app2}
\end{align}
Using the relationship between the standard and complementary error functions, this last integral can be written as the addition/subtraction of four separate integrals of the form $\int_{-\infty}^{\infty} \exp[-x^{2}]\mathrm{erfc}[Ax+B]$ or $\int_{-\infty}^{\infty} \exp[-x^{2}]\mathrm{erfc}[Ax+B]\mathrm{erfc}[Cx+D]$. Both of these can be solved by differentiating under the integral sign using the Leibniz integration rule, sometimes called the `Feynman integration trick'.\footnote{Again, with significant help from anonymous user \textsc{Przemo}'s solution to an  \href{https://math.stackexchange.com/questions/156047/integral-of-product-of-exponential-function-and-two-complementary-error-function}{`Integral of product of exponential function and two complementary error functions'} on \textsc{Mathematics Stack Exchange}.} The following identities for these integrals are
\begin{equation}
 \int_{-\infty}^{\infty} \exp[-x^{2}]\mathrm{erfc}[Ax+B] = \sqrt{\pi}\mathrm{erfc}\biggl[\frac{B}{\sqrt{1+A^{2}}}\biggl]
 \label{eq:appdef3}
\end{equation}
and
\begin{align}
 & \int_{-\infty}^{\infty} \exp[-x^{2}]\mathrm{erfc}[Ax+B]\mathrm{erfc}[Cx+D] = \notag \\
 & \sqrt{\pi}\biggl[1 + 4T(F_{1},F_{2})+4T(F_{3},F_{4}) + \frac{2}{\pi}\mathrm{tan}^{-1}(F_{5}) \notag \\ & \qquad -\frac{2}{\pi}\mathrm{tan}^{-1}(F_{2}) - \frac{2}{\pi}\mathrm{tan}^{-1}(F_{4}) + \mathrm{erf}\biggl(\frac{F_{1}}{\sqrt{2}}\biggl) + \mathrm{erf}\biggl(\frac{F_{3}}{\sqrt{2}}\biggl)\biggl] ~,
\end{align}
where $T(h, x)$ is Owen's T function and
\begin{align}
    & F_{1} = \frac{-\sqrt{2}B}{\sqrt{1+A^{2}}} \, , \, F_{2} = \frac{ABC-(1+A^{2})D}{B\sqrt{1+A^{2}+C^{2}}} \, , \, F_{3} = \frac{-\sqrt{2}D}{\sqrt{1+C^{2}}} \, , \\ & F_{4} = \frac{ACD-(1+C^{2})B}{D\sqrt{1+A^{2}+C^{2}}} \, , \, F_{5} = \frac{AC}{\sqrt{1+A^{2}+C^{2}}} ~.
\end{align}
Utilising these identities and comparing to the expression in Eq.~\ref{eq:app2} we arrive at our final result, a sum of elementary functions,
\begin{align}
    f_{n} &= \frac{1}{4}\mathrm{erf}\biggl(\frac{G^{\mathrm{max}}_{0}}{\sqrt{2}}\biggl) - \frac{1}{4}\mathrm{erf}\biggl(\frac{G^{\mathrm{min}}_{0}}{\sqrt{2}}\biggl) + \sum_{j=0}^{1}\biggl[T(G_{j}^{\mathrm{max}},H_{j}^{\mathrm{max}}) \notag \\ & -T(G_{j}^{\mathrm{min}},H_{j}^{\mathrm{min}}) + \frac{1}{2\pi}\mathrm{tan}^{-1}(H^{\mathrm{min}}_{j}) - \frac{1}{2\pi}\mathrm{tan}^{-1}(H^{\mathrm{max}}_{j})\biggl] ~,
    \label{eq:app3}
\end{align}
where
\begin{align}
    G^\mathrm{min(max)}_{0} &= -\sqrt{\frac{2}{1+G^{2}|\boldsymbol{\mathsf{C}}|}}R_{\mathrm{min(max)}} ~, \\
    G^\mathrm{min(max)}_{1} &= -\sqrt{\frac{\Psi_{ss}}{1+\delta|\boldsymbol{\mathsf{C}}|}}(s_{\mathrm{min}}-\bar{s}) ~, \\
    H^{\mathrm{min(max)}}_{0} &= \frac{G|\boldsymbol{\mathsf{C}}|\sqrt{\delta} - \sqrt{\frac{\Psi_{ss}}{2}}\frac{(s_{\mathrm{min}}-\bar{s})}{R_{\mathrm{min(max)}}}(1+G^{2}|\boldsymbol{\mathsf{C}}|)}{\sqrt{1+G^{2}|\boldsymbol{\mathsf{C}}|+\delta|\boldsymbol{\mathsf{C}}|}} ~, \\
    H^{\mathrm{min(max)}}_{1} &= \frac{G|\boldsymbol{\mathsf{C}}|\sqrt{\delta} - \sqrt{\frac{2}{\Psi_{ss}}}\frac{R_{\mathrm{min(max)}}}{(s_{\mathrm{min}}-\bar{s})}(1+\delta|\boldsymbol{\mathsf{C}}|)}{\sqrt{1+G^{2}|\boldsymbol{\mathsf{C}}|+\delta|\boldsymbol{\mathsf{C}}|}} ~, 
\end{align}
and $\delta$, $R_{\mathrm{min(max)}}$ and $G$ are as defined in Eqs.~\ref{eq:appdef1}-\ref{eq:appdef2}.

This result is a little complex but can still be coded up relatively easily. However, the benefit over numerically evaluating the 3D integral is substantial: it is both exact and can be evaluated much, much faster. This is especially true when the determinant and inverse of the covariance matrix are also computed in terms of the individual elements `manually' rather than relying on numerical matrix operations, which is trivial for a $3\times3$ matrix. As a comparison, numerically computing the determinant and inverse of the covariance matrix and using these as input to \textsc{scipy.tplquad} takes $\sim$1s to evaluate numerically for a single galaxy at one distance to reasonable precision. The analytic form above can be easily vectorised for many galaxies or values of $r
_{\mathrm{min(max)}}$ and then requires on average only $\sim$1$\mu$s per evaluation, a factor of $\sim$1,000,000 times faster.

\subsection{Derivation without magnitude limits}
\label{sec:appfn2}

When fitting the FP parameters, rather than the individual distances to each galaxy, we do not need to account for the magnitude limits of the data in the $f_{n}$ term as explained in Section~\ref{sec:FP}. In this case, the $f_{n}$ term becomes
\begin{multline}
    f_{n} = \frac{1}{(2\pi)^{3/2}|\boldsymbol{\mathsf{C}}|^{1/2}}\int_{-\infty}^{\infty}\int_{-\infty}^{\infty}\int_{s_{\mathrm{min}}}^{\infty}ds dr di \\ \exp\biggl\{-\frac{1}{2}(\boldsymbol{x}-\boldsymbol{\bar{x}})\boldsymbol{\mathsf{C}}^{-1}(\boldsymbol{x}-\boldsymbol{\bar{x}})^{T}\biggl\} ~,
\end{multline} 
which can also be evaluated in terms of elementary functions. Following the steps in the previous derivation, we see this integral can be solved by substituting $R_{\mathrm{min}} = -\infty$ and $R_{\mathrm{max}} = \infty$ into Eq.~\ref{eq:app2}, leaving
\begin{equation}
f_{n} = \frac{1}{2\sqrt{\pi}}\int_{-\infty}^{\infty} dx\, \exp\{-x^{2}\}\mathrm{erfc}\biggl\{\sqrt{\delta|\boldsymbol{\mathsf{C}}|}x + \sqrt{\frac{\Psi_{ss}}{2}}(s_{\mathrm{min}}-\bar{s})\biggl\} ~.
\end{equation}
We can again use the identity in Eq.~\ref{eq:appdef3} to write this as
\begin{equation}
f_{n} = \frac{1}{2}\mathrm{erfc}\biggl(-\frac{G_{1}^{\mathrm{min}}}{\sqrt{2}}\biggl) ~.
\end{equation}
This is equivalent to the 1D integral in Appendix~A of \cite{Magoulas2012}, but again is much faster to compute.

\section{Peculiar velocity error from incorrect distance ratio}
\label{sec:incorrect}

The correct relation between the log-distance ratio and physical size (effective radius) is
\begin{equation}
\eta = \log{R_e(z)} - \log{R_e(\bar{z})}
\label{eq:correct}
\end{equation}
where $z$ is the observed redshift and $\bar{z}$ is the cosmological redshift corresponding to the true comoving distance. These redshifts are related by 
\begin{equation}
(1+z) = (1+\bar{z})(1+z_p)
\label{eq:redshifts}
\end{equation}
where $z_p$ is the redshift corresponding to the peculiar velocity, $v_p = cz_p$. The incorrect relation used by \cite{Springob2014} is
\begin{align}
\eta' &= \log{R_e(z)} - \log{R_e(\bar{z})} + \log[{(1+z)/(1+\bar{z})}] \\
 &= \log{R_e(z)} - \log{R_e(\bar{z})}) + \log(1+z_{p}) \\
 &= \eta + \log(1+z_{p}) ~.
\label{eq:incorrect}
\end{align}
To proceed, we use the velocity estimator of \cite{Watkins2015}. This allows us to write 
\begin{equation}
\frac{v_{p}}{c} \approx \frac{z}{1+z}\ln(10)\eta ~,
\label{eq:watkins}
\end{equation}
which is accurate as long as $v_{p} \ll cz$. Substituting this into Eq.~\ref{eq:incorrect} and performing a Taylor expansion around $v_{p}=0$, we find
\begin{equation}
\eta' \approx \eta \biggl(1 - \frac{z}{1+z}\ln(10)\biggl)
\end{equation}
Hence, the relative error in the log-distance (and the peculiar velocity, according to Eq.~\ref{eq:watkins}) is given by
\begin{equation}
\frac{\eta v_p}{v_p} \equiv \frac{v_p^\prime - v_p}{v_p} 
\approx \frac{z}{1+z}\ln(10) ~.
\label{eq:zprelerr}
\end{equation}
Thus the relative error in the peculiar redshift (or peculiar velocity) from using the incorrect relation is approximately equal to two times the observed redshift.


\bsp	
\label{lastpage}
\end{document}